%% file: ms.tex
\documentclass[preprint,11pt]{aastex}
\usepackage{epsfig,natbib}
\usepackage{graphicx}
\usepackage{enumerate}
\usepackage{amsmath}
\usepackage{multirow}
\usepackage{bigdelim}
\usepackage{longtable}
\usepackage{threeparttable}
\usepackage{subfigure}
\usepackage{textcomp}
\usepackage{gensymb}
\usepackage{color}
\usepackage{morefloats}
\usepackage{xspace}
\usepackage[T1]{fontenc}
\usepackage{lscape}

\usepackage{hyperref}
\providecommand{\eprint}[1]{\href{http://arxiv.org/abs/#1}{#1}}

\providecommand{\adsurl}[1]{\href{#1}{ADS}}

\input{newcommand_gen.tex}
\makeatletter

\newcommand{\Rmnum}[1]{\expandafter\@slowromancap\romannumeral #1@}

\newcommand{\mytilde}{\raise.17ex\hbox{$\scriptstyle\mathtt{\sim}$}}
\newcommand{\specmatch}{\texttt{SpecMatch}\xspace}
\newcommand{\TERRA}{\texttt{TERRA}\xspace}

\newcommand{\Mstar}{\ensuremath{M_{\star}}\xspace}
\newcommand{\Rstar}{\ensuremath{R_{\star}}\xspace}

\newcommand{\Teff}{\ensuremath{T_{\mathrm{eff}}}\xspace}  
\newcommand{\logg}{\ensuremath{\log g}\xspace}

\newcommand{\Kepler}{\textit{Kepler}\xspace} 
\newcommand{\ktwo}{\textit{K2}\xspace}
\newcommand{\tess}{\textit{TESS}\xspace}

\newcommand{\nplanet}{{104}\xspace}  
\newcommand{\newplanet}{{64}\xspace}
\newcommand{\nfinalcand}{{63}\xspace} 
\newcommand{\nfp}{\ensuremath{30}\xspace}
\newcommand{\ntotcand}{{197}\xspace}
\newcommand{\nunval}{{28}\xspace}
\newcommand{\minfpp}{\ensuremath{0.01}\xspace}
\newcommand{\maxfpp}{\ensuremath{0.99}\xspace}
\newcommand{\fppall}{\ensuremath{15-30\%}\xspace}
\newcommand{\nsmall}{\ensuremath{37}\xspace} 
\newcommand{\nbright}{\ensuremath{15}\xspace} 

\renewcommand{\Re}{\ensuremath{R_{\oplus}}\xspace} 

\newcommand{\medrad}{\ensuremath{2.3\,\Re}\xspace} 
\newcommand{\medper}{\ensuremath{8.6}\,d\xspace} 
\newcommand{\medteff}{\ensuremath{5300}\,K\xspace} 
\newcommand{\medkp}{\ensuremath{12.7}\,mag\xspace}

\makeatother

\shortauthors{et all}
\shorttitle{K2 Planets in Fields 0--4}
\begin{document}
\pagenumbering{arabic}


\title{\ntotcand Candidates and \nplanet Validated Planets in K2's First Five Fields}
\author{
Ian J.\ M.\ Crossfield\altaffilmark{1,2},
David R.\ Ciardi\altaffilmark{3},
Erik A.\ Petigura\altaffilmark{4,5},
Evan Sinukoff\altaffilmark{6,7},
Joshua E.\ Schlieder\altaffilmark{8,9},
Andrew W.\ Howard\altaffilmark{6},
Charles A.\ Beichman\altaffilmark{3},
Howard Isaacson\altaffilmark{10},
Courtney D.\ Dressing\altaffilmark{4,2},
Jessie L.\ Christiansen\altaffilmark{3},
Benjamin J.\ Fulton\altaffilmark{6,11},
S\'ebastien L\'epine\altaffilmark{12},
Lauren Weiss\altaffilmark{10},
Lea Hirsch\altaffilmark{10},
John Livingston\altaffilmark{13},
Christoph Baranec\altaffilmark{14}, 
Nicholas M.\ Law\altaffilmark{15},
Reed Riddle\altaffilmark{16},
Carl Ziegler\altaffilmark{15},
Steve B. Howell,\altaffilmark{8},
Elliott Horch\altaffilmark{17},
Mark Everett\altaffilmark{18},
Johanna Teske\altaffilmark{19},
Arturo O.\ Martinez\altaffilmark{20},
Christian Obermeier\altaffilmark{21},
Bj\"orn Benneke\altaffilmark{4},
Nic Scott\altaffilmark{22},
Niall Deacon\altaffilmark{23},
Kimberly M.\ Aller\altaffilmark{6},
Brad M.\ S.\ Hansen\altaffilmark{24},
Luigi Mancini\altaffilmark{21},
Simona Ciceri\altaffilmark{25,21},
Rafael Brahm\altaffilmark{26,27},
Andr\'es Jord\'an\altaffilmark{26,27},
Heather A.\ Knutson\altaffilmark{4},
Thomas Henning\altaffilmark{21},
Micha\"el Bonnefoy\altaffilmark{28},
Michael C.\ Liu\altaffilmark{6},
Justin R.\ Crepp\altaffilmark{29},
Joshua Lothringer\altaffilmark{1},
Phil Hinz\altaffilmark{30},
Vanessa Bailey\altaffilmark{31,30},
Andrew Skemer\altaffilmark{32,28},
Denis Defrere\altaffilmark{33,28}}

\altaffiltext{1}{Lunar \& Planetary Laboratory, University of Arizona, 1629 E. University Blvd., Tucson, AZ, USA}
\altaffiltext{2}{NASA Sagan Fellow}
\altaffiltext{3}{NASA Exoplanet Science Institute, California Institute of Technology, Pasadena, CA, USA}
\altaffiltext{4}{Geological and Planetary Sciences, California Institute of Technology, Pasadena, CA, USA}
\altaffiltext{5}{Hubble Fellow}
\altaffiltext{6}{Institute for Astronomy, University of Hawai`i at M\={a}noa, Honolulu, HI, USA} 
\altaffiltext{7}{NSERC Postgraduate Research Fellow}
\altaffiltext{8}{NASA Ames Research Center, Moffett Field, CA, USA}
\altaffiltext{9}{NASA Postdoctoral Program Fellow}
\altaffiltext{10}{Astronomy Department, University of California, Berkeley, CA, USA}
\altaffiltext{11}{NSF Graduate Research Fellow}
\altaffiltext{12}{Department of Physics and Astronomy, Georgia State University, GA, USA}
\altaffiltext{13}{Department of Astronomy, The University of Tokyo, 7-3-1 Bunkyo-ku, Tokyo 113-0033, Japan}
\altaffiltext{14}{Institute for Astronomy, University of Hawai`i at M\={a}noa, Hilo, HI, USA}
\altaffiltext{15}{Department of Physics and Astronomy, University of North Carolina at Chapel Hill, Chapel Hill, NC, USA}
\altaffiltext{16}{Division of Physics, Mathematics, and Astronomy, California Institute of Technology, Pasadena, CA, USA}
\altaffiltext{17}{Department of Physics, Southern Connecticut State University, New Haven, CT, USA}
\altaffiltext{18}{National Optical Astronomy Observatory, Tucson, AZ, USA}
\altaffiltext{19}{Carnegie Department of Terrestrial Magnetism, Washington, DC, USA}
\altaffiltext{20}{Department of Astronomy, San Diego State University, San Diego, CA, USA}
\altaffiltext{21}{Max Planck Institut f\"ur Astronomie, Heidelberg, Germany}
\altaffiltext{22}{Sydney Institute of Astronomy, The University of Sydney, Redfern, Australia}
\altaffiltext{23}{Centre for Astrophysics Research,University of Hertfordshire, UK}
\altaffiltext{24}{Department of Physics \& Astronomy and Institute of Geophysics \& Planetary Physics, University of California Los Angeles, Los Angeles, CA, USA}
\altaffiltext{25}{Department of Astronomy, Stockholm University, SE-106 91 Stockholm, Sweden}
\altaffiltext{26}{Millennium Institute of Astrophysics, Av.\ Vicu\~na Mackenna 4860, 7820436 Macul, Santiago, Chile}
\altaffiltext{27}{Instituto de Astrof\'isica, Facultad de F\'isica, Pontificia Universidad Cat\'olica de Chile, Av.\ Vicu\~na Mackenna 4860, 7820436 Macul, Santiago, Chile}
\altaffiltext{28}{Univ. Grenoble Alpes, IPAG, 38000, Grenoble, France; CNRS, IPAG, 38000, Grenoble, France}
\altaffiltext{29}{Department of Physics, University of Notre Dame, 225 Nieuwland Science Hall, Notre Dame, IN, USA}
\altaffiltext{30}{Steward Observatory, The University of Arizona, Tucson, AZ, USA}
\altaffiltext{31}{Kavli Institute for Particle Astrophysics and Cosmology, Stanford University, Stanford, CA, USA}
\altaffiltext{32}{Department of Astronomy, University of California, Santa Cruz, Santa Cruz, CA, USA}
\altaffiltext{33}{Departement d'Astrophysique, Geophysique et Oceanographie, Universite de Liege, 4000 Sart Tilman, Belgium}

\begin{abstract}

  We present \ntotcand planet candidates discovered using data from
  the first year of the NASA \ktwo mission (Campaigns 0--4), along
  with the results of an intensive program of photometric analyses,
  stellar spectroscopy, high-resolution imaging, and statistical
  validation.  We distill these candidates into sets of \nplanet
  validated planets (57 in multi-planet systems), \nfp false
  positives, and \nfinalcand remaining candidates. Our validated
  systems span a range of properties, with median values of
  $R_P$\,=\,\medrad, $P$\,=\,\medper, \Teff\,=\,\medteff, and {\em
    Kp}\,=\,\medkp.  Stellar spectroscopy provides precise stellar and
  planetary parameters for most of these systems.  We show that \ktwo
  has increased by 30\% the number of small planets known to orbit
  moderately bright stars (1--4\,$R_\oplus$, $Kp$\,=\,9--13~mag). Of
  particular interest are \nsmall planets smaller than 2\,$R_\oplus$,
  \nbright orbiting stars brighter than $Kp$\,=\,11.5~mag, five receiving
  Earth-like irradiation levels, and several multi-planet systems ---
  including four planets orbiting the M dwarf K2-72 near mean-motion
  resonances.  By quantifying the likelihood that each candidate is a
  planet we demonstrate that our candidate sample has an overall false
  positive rate of \fppall, with rates substantially lower for small
  candidates ($<2R_\oplus$) and larger for candidates with radii
  $>8R_\oplus$ and/or with $P<3$\,d.  Extrapolation of the current
  planetary yield suggests that \ktwo will discover between
  $500\,-\,1000$ planets in its planned four-year mission --- assuming
  sufficient follow-up resources are available.  Efficient observing
  and analysis, together with an organized and coherent follow-up
  strategy, is essential to maximize the efficacy of planet-validation
  efforts for \ktwo, \tess, and future large-scale surveys.
\end{abstract}


\section{Introduction}
\label{sec:intro}

Planets that transit their host stars offer unique opportunities to
characterize planetary masses, radii, and densities; atmospheric
composition, circulation, and chemistry; dynamical interactions in
multi-planet systems; and orbital alignments and evolution, to name
just a few aspects of interest.  Transiting planets are also the most
common type of exoplanet known, thanks in large part to NASA's \Kepler
spacecraft. Data from \Kepler's initial four-year survey revealed over
4000 candidate exoplanets and many confirmed and validated
planets\footnote{We distinguish ``confirmed'' systems (with measured
  masses) from ``validated'' systems (whose planetary nature has been
  statistically demonstrated, e.g. with false positive probability
  $<$\,1\% ).}  \citep[e.g.,][]{coughlin:2016,morton:2016}. A
majority of all exoplanets known today were discovered by \Kepler.
After the spacecraft's loss of a second reaction wheel in 2014, the
mission was renamed \ktwo and embarked on a new survey of the ecliptic
plane, divided into campaigns of roughly 80 days each
\citep{howell:2014}.  In terms of survey area, temporal coverage, and
data release strategy, \ktwo provides a natural transition from
\Kepler to the \tess mission \citep{ricker:2014}.  \Kepler observed
1/400$^{th}$ of the sky for four years (initially with a default
proprietary period), while \tess will observe nearly the entire sky
for $\ge27$ days\footnote{Smaller fractions of the sky will be
  observed for up to 351\,d.}, with no default proprietary period.
 
In its brief history \ktwo has already made many new discoveries.  The
mission's data have helped to reveal oscillations in variable stars
\citep{angus:2016} and discovered eclipsing binaries
\citep{lacourse:2015,armstrong:2016,david:2016a}, supernovae
\citep{zenteno:2015}, large numbers of planet candidates
\citep{foremanmackey:2015,vanderburg:2016,adams:2016}, and a growing
sample of validated and/or confirmed planets
\citep[e.g.,][]{vanderburg:2014,crossfield:2015a,sanchis-ojeda:2015,huang:2015,montet:2015,sinukoff:2016}.
Here, we report our identification and follow-up observations of
\ntotcand candidate planets using \ktwo data.  Using all available
observations and a robust statistical framework, we validate \nplanet
of these as true, {\em bona fide} planets and for the remaining
systems discriminate between obvious false positives and a remaining
subset of plausible candidates suitable for further follow-up.

In Sec.~\ref{sec:phot} we review our target sample, photometry and
transit search, and initial target vetting.  Sec.~\ref{sec:obs}
describes our supporting ground-based observations (stellar
spectroscopy and high-resolution imaging), while Sec.~\ref{sec:stars}
describes our derivation of stellar parameters. These are followed by
our intensive transit light curve analysis in Sec.~\ref{sec:tlc}, the
assessment of false positive probabilities for our candidates in
Sec.~\ref{sec:fpp}, and a discussion of the results, interesting
trends, and noteworthy individual systems in
Sec.~\ref{sec:discussion}. Finally, we conclude and summarize in
Sec.~\ref{sec:conclusion}.

\section{\ktwo Targets and Photometry}
\label{sec:phot}

\subsection{Target Selection}


In the analysis that follows we use data from all \ktwo targets (not
just those in our own General Observer proposals\footnote{K2
  Programs 79, 120, 1002, 1036, 2104, 2106, 2107, 3104, 3106, 3107,
  4011, 4033}).  \cite{huber:2016} present the full distribution of
stellar types observed by \ktwo. For completeness we describe here our
target selection strategy, which has successfully proposed for
thousands of FGK and M dwarfs through two parallel efforts.

We select our FGK stellar sample from the all-sky {\it TESS} Dwarf
Catalog \citep[TDC;][]{stassun:2014}. The TDC consists of 3 million
F5--M5 candidate stars selected from 2MASS and cross-matched with the
NOMAD, Tycho-2, Hipparcos, APASS, and UCAC4 catalogs to obtain
photometric colors, proper motions, and parallaxes.  We remove giant
stars based on reduced proper motion vs.\ $J-H$ color
\citep[see][]{cameron:2007b}, and generate a magnitude-limited dwarf
star sample from the merged TDC/EPIC by requiring {\em
  Kp}\,$<$\,14~mag for these FGK stars.  We impose an anti-crowding
criterion and remove all targets with a second star in EPIC
\citep[complete down to {\em Kp}\,$\sim$\,19 mag;][]{huber:2016}
within 4 arcsec (approximately the Kepler pixel size).  This last
criterion removes $<1$\% of the FGK stars in our proposed samples,
improves catalog reliability by reducing false positives, and
simplifies subsequent vetting and Doppler follow-up.

We draw our late-type (K and M dwarf) stellar sample primarily from
the SUPERBLINK proper motion database \citep[SB,][]{lepine:2005} and
the PanSTARRS-1 survey \citep[PS1,][]{kaiser:2002}.  We use a
combination of reduced proper motion, optical/NIR color cuts, and/or
SED fitting to capture the majority of M dwarfs ($>$85\%) within 100
pc with little contamination from distant giants.  In some \ktwo
campaigns we supplement our initial database using SDSS, PS1, and/or
other photometry to identify additional targets with smaller proper
motions \citep[following ][]{aller:2013}.  We estimate approximate
spectral types (SpTs) using tabulated photometric relations
\citep{kraus:2007,pecaut:2013,rodriguez:2013} and convert SpTs into
stellar radii ($R_*$) based on interferometric studies
\citep{boyajian:2012b}.  Our exact selection criteria for K and M
dwarfs have evolved with time, but we typically prioritize this
low-temperature stellar sample by requiring S/N\,$\gtrsim8$ for a
single transit of an Earth-sized planet, assuming the demonstrated
photometric precision of \ktwo. We additionally set a magnitude limit
of Kp\,$<$\,16.5~mag on this late-type dwarf sample to allow feasible
spectroscopic characterization.

\subsection{Time-Series Photometry}

Our team's photometric pipeline \citep[described by,
  e.g.,][]{crossfield:2015a,petigura:2015a} builds on the approach
originally outlined by \cite{vanderburg:2014}.  We extract time-series
photometry from the target pixel files provided by the project using
circular, stationary, soft-edged apertures. During \ktwo operations,
solar radiation pressure torques the spacecraft, causing it to roll
around the boresight. This motion causes a typical target star to
drift across the CCD by $\sim$1 pixel every $\sim$6 hours. This motion
of stars across the CCD, when combined with inter- and intra-pixel
sensitivity variations and aperture losses, results in significant
changes in our aperture photometry.

We remove these stellar brightness variations that correlate with
spacecraft orientation by first solving for the roll angle between
each frame and an arbitrary reference frame using roughly 100 stars of
Kp$\sim$12~mag on an arbitrary output channel. Then, we model the
time- and roll-dependent brightness variations using a Gaussian
process with a squared-exponential kernel. We apply apertures with radii ranging from 1--7 pixels and
select the aperture that minimizes the residual noise in the corrected
light curve (computed on three-hour timescales). This minimization
balances two competing effects: larger apertures yield smaller
systematic errors (because aperture losses are smaller) while smaller
apertures incur less background noise. All our processed light curves
are available for download at the NExScI ExoFOP
website\footnote{\url{https://exofop.ipac.caltech.edu}}.

\subsection{Identifying Transit-Like Signals}
\label{sec:vetting}
We search our calibrated photometry for planetary transits using the
\TERRA algorithm \citep{petigura:2013b}.  After running \TERRA, we
flag stars with putative transits having signal to noise (S/N) $>$ 12
as threshold-crossing events (TCEs) for visual inspection. Below this
level, transit signals surely persist but TCEs become dominated by
spurious detections. Residual outliers in our photometry prevent us
from identifying large numbers of candidates at lower S/N. In order to
reduce the number of spurious detections we require that TCEs have
orbital periods $P\ge1$\,d, and that they also show three
transits. This last criterion sets an upper bound to the longest
period detectable in our survey at half the campaign baseline or
$\sim$37\,d\footnote{The handful of candidates with $P>37$\,d were
  found by visual inspection.}.  Thus many longer-period planets
likely remain to be found in these data sets, in a manner analogous to
the discovery of HIP-116454b in \ktwo's initial engineering run
\citep{vanderburg:2015} and additional single-transit candidates
identified in Campaigns 1--3 \citep{osborn:2016}.

\begin{figure}[ht!]
\begin{center}
\includegraphics[width=3.5in]{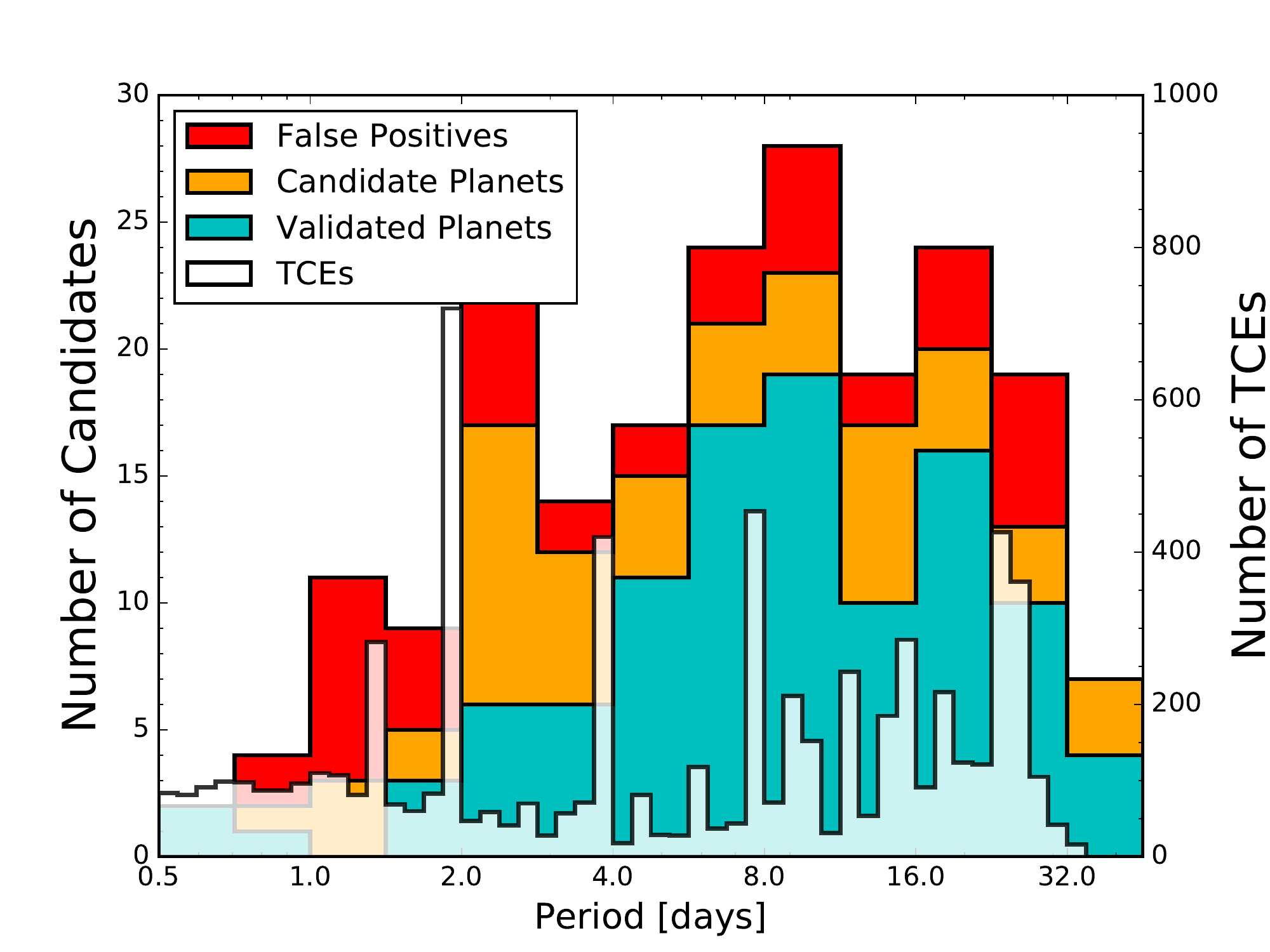}
\caption{\label{fig:period_dist} Distribution of orbital periods of transit-like
  signals identified in our analysis. The pale, narrow-binned histogram (axis at
  right) indicates the Threshold-Crossing Events (TCEs) identified by
  \TERRA in our initial transit search (see Sec.~\ref{sec:phot}).  The
  coarser histograms (axis at left) indicate the cumulative distributions of
  \nplanet validated planets (blue-green; FPP$<$\minfpp), \nfp false positive systems
  (red; FPP$>$\maxfpp), and \nfinalcand candidates of indeterminate status
  (orange). }
\end{center}
\end{figure}

In our analysis, each campaign yields roughly 1000 TCEs. The
distribution of their orbital periods, shown in Fig.~\ref{fig:period_dist},
reveals discrete peaks at $P$=1.5, 2, 4, 8, and 16
days. These sharp peaks likely correspond to the 6\,hr periodicity of
small-scale maneuvering tweaks to rebalance solar pressure and/or to
the 48\,hr periodicity of \ktwo's reaction-wheel momentum dumps
\citep{vancleve:2015}. Both these effects could induce correlated photometric
jitter on integer multiples of this timescale. We also see a smoother increase
in TCEs toward longer periods ($P\gtrsim16$\,d) that our manual
vetting (described below) shows to correspond to an increasing false
positive rate for TCEs showing just 3--5 transit-like events.

\begin{figure}[ht!]
\begin{center}
\includegraphics[width=5.5in]{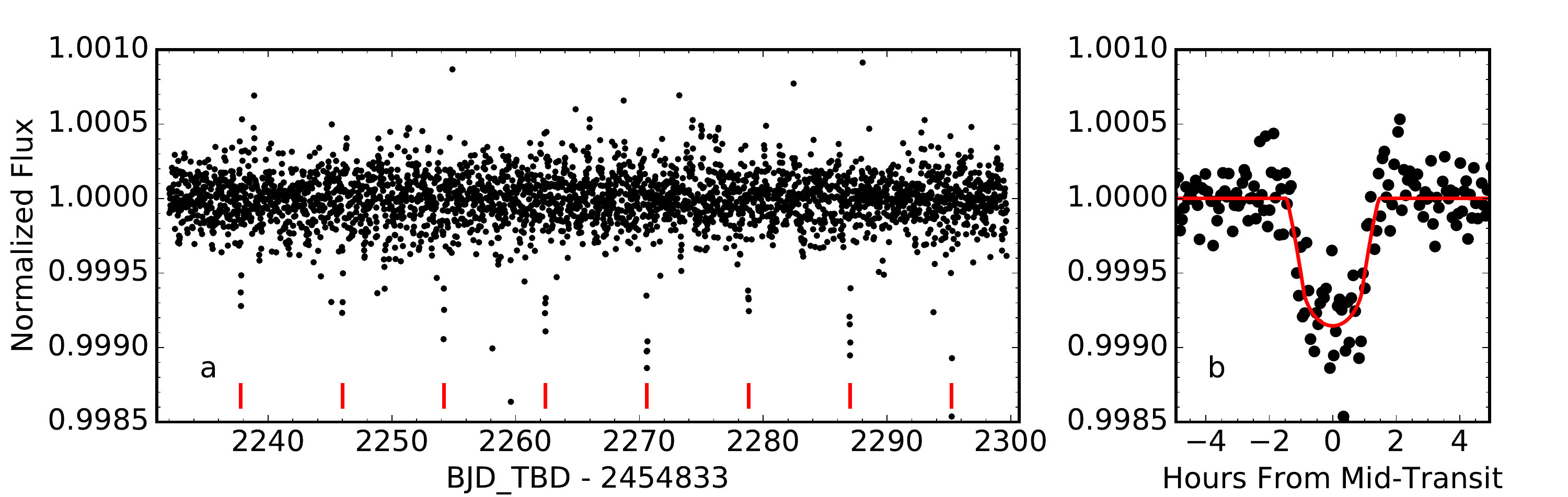}
\caption{\label{fig:demo_lightcurve} Example light curve of
  K2-77 (EPIC~210363145), which hosts one validated planet: (a) during all of
  Campaign 4, with individual transit times indicated, and (b)
  phase-folded, with the best-fit transit model overplotted in
  red. The transit parameters for all candidates are listed in
  Table~\ref{tab:results}.}
\end{center}
\end{figure}

In each campaign, our manual vetting process begins with these TCEs
and results in well-defined lists of astrophysical variables,
including robust planet candidates for further follow-up and
validation.  \TERRA produces a set of diagnostics for every TCE with a
detection above our $S/N$ limit, which we use to determine whether the
event was likely caused by a candidate planet, eclipsing binary,
periodic variable, or noise. The diagnostics include a summary of
basic fit parameters and a suite of diagnostic plots to visualize the
nature of the TCE. These plots include the \TERRA periodogram, a
normalized phase-folded light curve with best fit model, the light
curve phased to 180$^{\circ}$ to look for eclipses or misidentified
periods, the most probable secondary eclipse identified at any phase,
and an auto-correlation function.  When vetting the user flags
each TCE as an object of interest or not, where objects of interest
can be either candidate planets, eclipsing binaries, or variable
stars.  We elevate any TCE showing no obvious warning signs to the
status of ``planet candidate,'' i.e.\ an event that is almost surely
astrophysical in nature, possibly a transiting planet, and not
obviously a false positive scenario like a background eclipsing
binary. We quantify the false positive probabilities of all our
candidates in Sec.~\ref{sec:fpp}.  Fig.~\ref{fig:demo_lightcurve}
shows an example of a \TERRA-derived light curve for a typical
candidate.

Once we identify a candidate, we re-run \TERRA to search for
additional planets in that system as described by
\cite{sinukoff:2016}. In brief, we mask out the photometry associated
with transits of the previously identified candidate and run \TERRA
again to look for additional box-shaped signals. We repeat this
process until no candidates are identified with S/N $>$ 8 or the
number of candidates exceeds five. We typically find $<10$
multi-candidate systems per campaign, with a maximum of four planets
detected per star.

\begin{figure}[ht!]
\begin{center}
\includegraphics[width=5.5in]{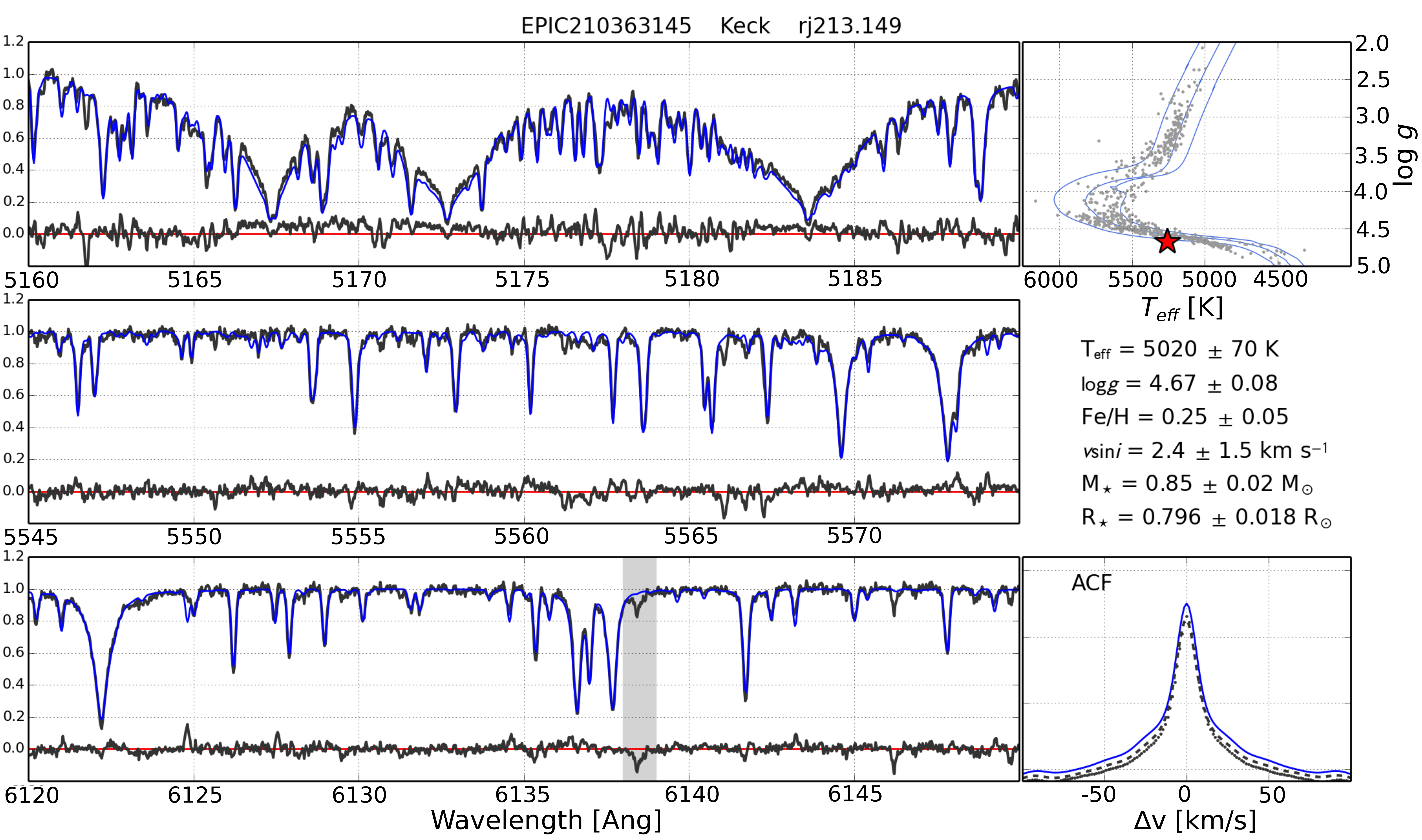}
\caption{\label{fig:demo_specmatch} Example Keck/HIRES stellar
  spectrum (blue), template match (black), and derived parameters for
  K2-77 (EPIC~210363145). The star has low $v \sin i$, moderate \Teff, and
  shows no evidence for additional stellar companions in the
  spectroscopic autocorrelation function (ACF). The upper-right panel plots 
  the derived stellar parameters against the parameters of the
  \specmatch\ template stars.  Stellar parameters for all targets are
  listed in Table~\ref{tab:stars}, and results of ACF analyses are in
  Table~\ref{tab:reamatch}.}
\end{center}
\end{figure}

\begin{figure}[ht!]
\begin{center}
\includegraphics[width=3.5in,angle=180]{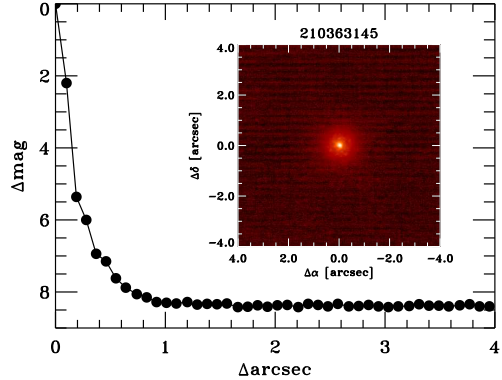}
\caption{\label{fig:demo_ao} Example constraints on any additional,
  nearby stars around K2-77 (EPIC~210363145) from Keck/NIRC2 K-band adaptive
  optics imaging. For this target, no companions were detected above
  the plotted contrast limits. Detected stellar companions around all
  observed candidates are listed in Table~\ref{tab:aocomp}. }
\end{center}
\end{figure}



\section{Supporting Observations}
\label{sec:obs}
\subsection{High-Resolution Spectroscopy: Observations}
\label{sec:spectra}

\subsubsection{Keck/HIRES}
We obtained high resolution optical spectra of 83 planet candidate
hosts using the HIRES echelle spectrometer \citep{vogt:1994} on the
10\,m Keck I telescope.  These spectra were collected using the
standard procedures of the California Planet Search
\citep[CPS;][]{howard:2010}.  We used the ``C2'' decker ($0\farcs87$
$\times$ 14\arcsec{} slit), which is long enough to simultaneously
measure the spectra of the target star and the sky background with
spectral resolution $R$\,=\,55,000.  The sky was subtracted from each
stellar spectrum.  We used the HIRES exposure meter to automatically
terminate each exposure once the desired S/N was reached, typically
after 1--20 min.  For stars with $V<$ 13.0~mag, exposure levels were
set to achieve S/N=45 per pixel at 550 nm.  Exposures of fainter stars
were terminated at S/N=32 per pixel---enough to derive stellar
parameters while keeping exposure times reasonable.  For 
stars that were part of subsequent Doppler campaigns, we measured
additional HIRES spectra with higher S/N. These RV measurements will
be the subject of a series of forthcoming papers.

\subsubsection{APF/Levy}

We obtained spectra of 27 candidate host stars using the Levy
high-resolution optical spectrograph mounted at the Automated Planet
Finder (APF). Each spectrum covers a continuous wavelength range from
374\,nm to 970\,nm. We observed the stars using either the
2''$\times$8'' slit for a spectral resolution of $R\approx80,000$ or,
to minimize sky contamination, the 1''$\times$3'' slit for a spectral
resolution of $R\approx100,000$. We initially observed all bright
targets using the 2''$\times$8'' slit to maximize S/N but soon noticed
that sky contamination was a serious problem on nights with a full or
gibbous moon. All APF spectra collected after 21 May 2015 were
observed using the 1$\times$3'' decker. In all cases, we collected
three consecutive exposures and combined the extracted 1D spectra
using a sigma-clipped mean to reject cosmic rays. All targets were
observed at just a single epoch. The final S/N of the combined spectra
ranges from roughly 25 to 50 per pixel.

\subsubsection{MPG 2.2\,m/FEROS}
We obtained spectra of a small number of candidate stellar hosts using
the FEROS fiber-fed echelle spectrograph \citep{kaufer:1998} at the
2.2\,m MPG telescope. Each spectrum covers a continuous wavelength
range from 350\,nm to 920\,nm with an average resolution of
R$\sim$48,000.  Our FEROS exposure times were chosen according to the
brightness of each target and ranged from 10--30\,min. Simultaneously
with the science images we acquired spectra of a ThAr lamp for
wavelength calibration.

The FEROS data are processed through a dedicated pipeline built from a
modular code (CERES, Brahm et al. in prep) designed to reduce, extract
and analyze data from different echelle spectrographs in an automated,
homogeneous and robust way.  This pipeline is similar to the
calibration and optimal extraction approach described by
\cite{jordan:2014}.  We compute a global wavelength solution from the
calibration ThAr image by fitting a Chebyshev polynomial as function
of the pixel position and echelle order number. The instrumental
velocity drifts during the night are computed using the the extracted
spectra of the ThAr lamp acquired during the science observations with
the reference fiber.  The barycentric correction is performed using
the JPLephem package. Radial velocities (RVs) and bisector spans are
determined by cross-correlating the continuum-normalized stellar
spectrum with a binary mask derived from a G2 dwarf's spectrum
\citep[for more details see, e.g.,][]{baranne:1979,queloz:1995}. We
normalize the stellar continuum to minimize the systematic errors that
would be induced in the derived velocity by differences in spectral
slope caused by  different reddening or stellar type.

\begin{deluxetable}{l l}
\tabletypesize{\footnotesize}
\tablecaption{FEROS Follow-up Observations}
\tablehead{EPIC     & Observation Note}
 \startdata
201176672 & Multiple peaks in CCF; likely stellar blend.  \\
201270176 & Multiple peaks in CCF; likely stellar blend. \\
202088212 & Multiple peaks in CCF; likely stellar blend. \\
203929178 & Multiple peaks in CCF; likely stellar blend.  \\
204873331 & Multiple peaks in CCF; likely stellar blend.  \\
203485624 & Very broad CCF peak, $v \sin i>50$ km~s$^{-1}$. \\
205148699 & Single-peaked CCF, phased RV variations of $\pm28$~km~s$^{-1}$. \\
201626686 & Single-peaked CCF, unphased  RV jitter of $\pm$50~m~s$^{-1}$. \\
203771098 & Single-peaked CCF, RV variations $<$20~m~s$^{-1}$.  \\
201505350 & Single-peaked CCF, $\sim$20~m~s$^{-1}$ RV variation between two epochs.  \\
201862715 & Single-peaked CCF.  \\
\enddata
\label{tab:feros}
\end{deluxetable}



\subsection{High-Resolution Spectroscopy: Methods and Results}
\label{sec:hrsresults}
As part of our false positive analysis (described in
Sec.~\ref{sec:fpp}), we use our high-resolution Keck/HIRES spectra to
search for additional spectral lines in the stellar spectra. This
method is sensitive to secondary stars that lie within 0.4'' of the
primary star (one half of the slit width) and that are as faint as 1\%
of the apparent brightness of the primary star \citep{kolbl:2015}. The
approach therefore complements the AO and speckle imaging described in
Sec.~\ref{sec:ao} \citep{teske:2015,ciardi:2015}.

The search for secondary lines in the HIRES spectra begins with a
match of the primary spectrum to a catalog of nearby, slowly-rotating,
FGKM stars from the CPS.  The best match from the catalog is
identified, subtracted from the primary spectrum, and the residuals
are then searched (using the same catalog) to identify any fainter
second spectrum.  This method is insensitive to companion stars with
velocity offsets of $\lesssim$10~km~s$^{-1}$, in which cases multiple
stellar lines would be blended too closely together. This method is
optimized for slowly rotating FGKM stars, so stars earlier than F and
those with $v \sin i>$10~km~s$^{-1}$ are more difficult to detect due
to their having fewer and/or broader spectral lines.  The technique is
less sensitive for stars with $\Teff\lesssim$3500~K due to the small
number of such stars in the CPS catalog.  The derived constraints for
all targets are listed in Table~\ref{tab:reamatch}, and we use them in
our false positive analysis described in Sec.~\ref{sec:fpp}.
Fig.~\ref{fig:demo_specmatch} shows an example of a Keck/HIRES
spectrum, together with the secondary line search results and derived
stellar parameters (see Sec.~\ref{sec:stars}).

We performed a similar analysis for the subset of stars observed by
the FEROS spectrograph. Table~\ref{tab:feros} lists these stars, most
of which host candidate hot Jupiters. Three show obvious signs of
multiple peaks in the stellar cross-correlation, indicating these
sources are blends of multiple stars; a fourth shows an extremely high
rotational velocity. As described in Sec.~\ref{sec:fpp}, we find false
positive probabilities (FPPs) of $>$50\% for all four of these
systems, indicating that most are likely false positives and
low-priority targets for future follow-up.

By obtaining FEROS spectra at multiple epochs, we detect RV variations
from EPIC~205148699 in phase with the transit signal and with
semi-amplitude $K\sim28$~km~s$^{-1}$, indicating that this system is
an eclipsing stellar binary.  For EPIC~201626686, 11 RV measurements
over 40 days reveal variations at the level of
$\pm$50~m~s$^{-1}$. Since these variations are not in phase with the
orbital period of the detected transits, we do not consider this
system to be a false positive.  Finally, multiple RV measurements also
set an upper limit on the RV variations of K2-24 (EPIC~203771098) of
$<20$~m~s$^{-1}$ \citep[consistent with the analysis of
][]{petigura:2016a}. Our analysis in Sec.~\ref{sec:fpp} ultimately
finds FPP$<$\minfpp for all three of these systems, indicating that
these are validated planets.

Single-epoch FEROS observations reveal that both K2-19
(EPIC~201505350) and EPIC 201862715 are single-lined dwarf stars,
consistent with our validation of the former. He latter has a close
stellar companion that prevents us from validating the system (see
Sec.~\ref{sec:fpp}), but radial velocity measurements have confirmed
the planet as WASP-85Ab.  A second observation of K2-19 taken three
days later shows an RV variation of $\sim$20~m~s$^{-1}$, roughly
consistent with the RV signal reported by \cite{barros:2015}.

\subsection{High-resolution Imaging}
\label{sec:ao}

\subsubsection{Observations}


We obtained high-resolution imaging (HRI) for 164 of our candidate
systems. Our primary instrument for this work was NIRC2 at the 10\,m Keck~II
telescope, with which we observed 110 systems. Most were observed in
Natural Guide Star (NGS) mode, though we used Laser Guide Star (LGS)
mode for a subset of targets orbiting fainter stars.  As part of multi-semester program
GN-2015B-LP-5 (PI Crossfield) at Gemini Observatory, we observed 40
systems with the NIRI camera \citep{hodapp:2003} in K band using NGS
or LGS modes.  We also observed 33 stars with PHARO/PALM-3000
\citep{hayward:2001, dekany:2013} at the 5\,m Hale Telescope and 14
systems with LMIRCam at LBT \citep{leisenring:2012}, all at K band.
We observed 39 stars at visible wavelengths using the automated
Robo-AO laser adaptive optics system at the Palomar 1.5\,m telescope
\citep{baranec:2013,baranec:2014}. These data were acquired and
reduced separately using the standard Robo-AO procedures outlined by
\cite{law:2014}.

We acquired the data from all our large-aperture AO
observations (NIRC2, NIRI, LMIRCam, PHARO) in a consistent manner.  We
observed at up to nine dither positions, using integration times short
enough to avoid saturation (typically $\le$60\,s). We use the dithered
images to remove sky background and dark current, and then align,
flat-field, and stack the individual images.

Through our Long-Term Gemini program we also acquired high-resolution
speckle imaging of 32 systems in narrow band filters centered at
692\,nm and 880\,nm using the DSSI camera
\citep{horch:2009,horch:2012b} at the Gemini North telescope. The DSSI
observing procedure is typically to center the target star in the
field, set up guiding, and take data using 60\,ms exposures.  The
total integration time varies by target brightness and observing
conditions. We measure background sensitivity in a series of
concentric annuli around the target star.  The innermost data point
represents the telescope diffraction limit, within which we set our
sensitivity to zero.  After measuring our sensitivity across the DSSI
field of view, we interpolate through the measurements using a cubic
spline to produce a smooth sensitivity curve.

\subsubsection{Contrast \& Stellar Companions}
\begin{figure}[ht!]
\begin{center}
\includegraphics[width=3.5in]{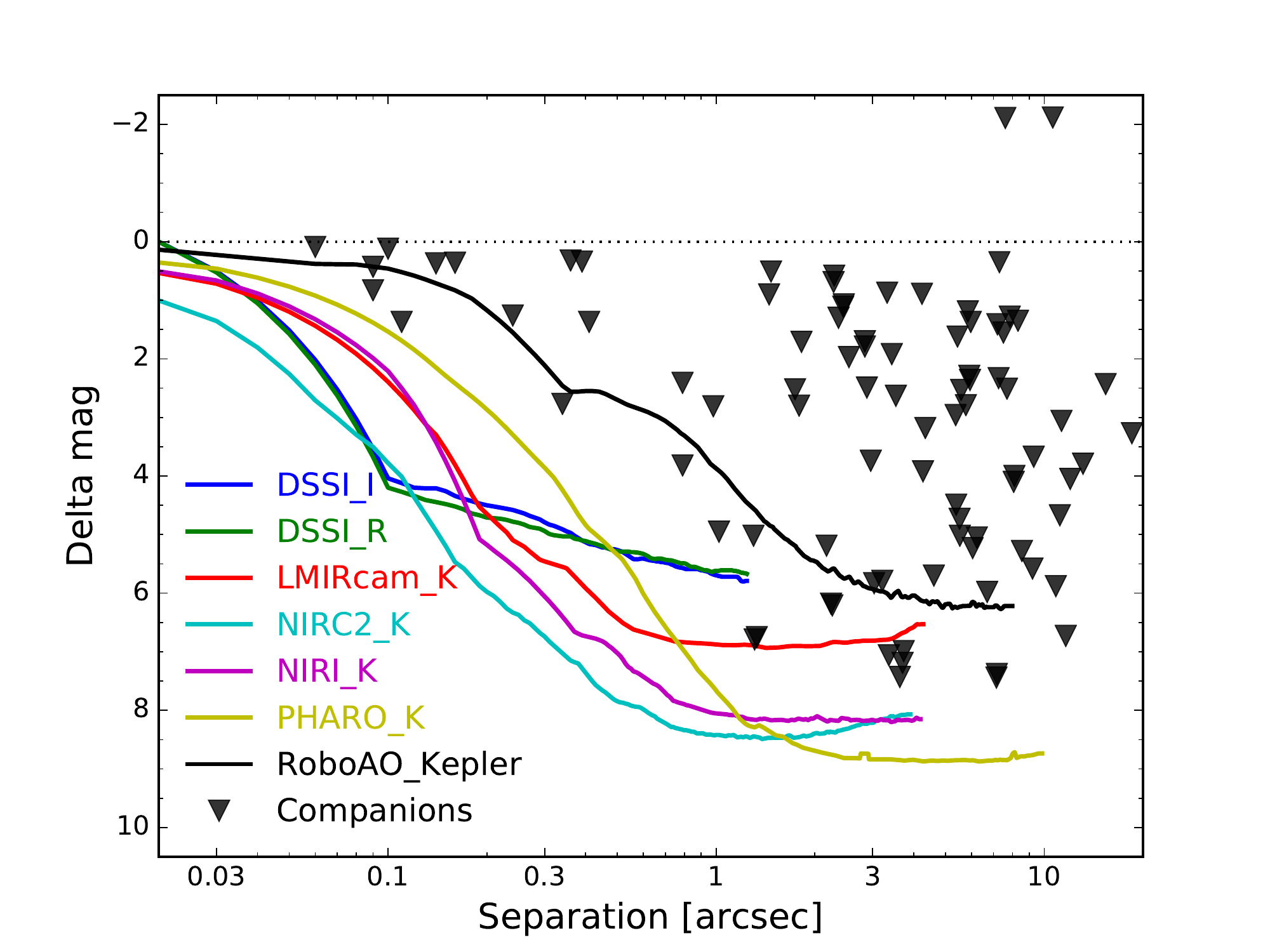}
\caption{\label{fig:aocomp} Stellar companions (triangles) detected near our \ktwo
  candidate systems and the median contrast achieved with
  each listed instrument and filter (solid curves).  As described in
  Sec.~\ref{sec:ao}, these detected magnitude differences set upper
  limits on the maximum amount of blending possible within the \Kepler
  bandpass. Parameters of these nearby stars are listed in
  Table~\ref{tab:aocomp}.  }
\end{center}
\end{figure}

We estimate the sensitivity of all our HRI data by injecting fake
sources 
into the final combined images with separations at integral multiples
of the central source's FWHM \citep[see
  e.g.][]{adams:2012,ziegler:2016}.  Fig.~\ref{fig:demo_ao} shows an
example of Keck/NIRC2 NGS image and the resulting 5$\sigma$ contrast
curve.  The median contrast curves achieved by each HRI instrument are
shown in Fig.~\ref{fig:aocomp} together with all detected stellar
companions.  The companions are also listed in Table~\ref{tab:aocomp}.
Contrast curves for each individual system are included as an
electronic supplement, and on the ExoFOP website.  In addition,
Table~\ref{tab:hri} includes the total integration times and filters
used for all candidates observed in our follow-up efforts.

The contrast curves are plotted in the band of observations, which
ranges from optical wavelengths (DSSI; Robo-AO) to K band
(large-aperture AO systems).  These in-band magnitude differences set
upper limits on the maximum amount of blending possible within the
\Kepler bandpass.  If the companion has the same color as the primary,
then the measured $\Delta$mag is indeed the $\Delta${\em Kp}. If the
companion is redder, then the {\em Kp}-band flux ratio is even
smaller.
All detected sources are included in Table~\ref{tab:aocomp}, even
though some lie outside of our photometric apertures.  In these cases
the detected companion has little or no impact on the transit
parameters and false positive probabilities derived below.  We discuss
such considerations more thoroughly in Sec.~\ref{sec:dilution}. 

\section{Stellar Parameters}
\label{sec:stars}
Stellar parameters are needed to convert the physical properties measured by our transit photometry into useful
planetary parameters such as radius ($R_P$) and incident irradiation
($S_{inc}$).  We use several complementary techniques to infer stellar
parameters for our entire sample.

For all stars with Keck/HIRES and/or APF/Levy spectra, we attempt to
estimate effective temperatures, surface gravities, metallicities, and
rotational velocities using \texttt{SpecMatch}
\citep{petigura:2015phd}.  \texttt{SpecMatch} fits a high-resolution
optical spectrum to an interpolated library of model spectra from
\citet{coelho:2005}, which closely match the spectra of
well-characterized stars in this temperature range.  Uncertainties on
\Teff, \ensuremath{\log g}, and
\ensuremath{[\mbox{Fe}/\mbox{H}]~} from HIRES spectra are 60 K, 0.08--0.10 dex, and 0.04
dex, respectively \citep[][]{petigura:2015phd}. 
Experience shows that \texttt{SpecMatch} is limited to stars with
$\Teff\sim$4700--6500~K and $v \sin i \lesssim 30$~km~s$^{-1}$.

The \texttt{SpecMatch} pipeline used to analyze the APF data is identical to
the Keck \texttt{SpecMatch} pipeline except that we employ the
differential-evolution Markov Chain Monte Carlo
\citep[DE-MCMC;][]{terbraak:2006} fitting engine from ExoPy
\citep{fulton:2013} instead of $\chi^2$ minimization. The APF
\texttt{SpecMatch} pipeline was empirically calibrated to produce consistent
stellar parameters for stars that were observed at both Keck and APF
by fitting and subtracting a 3-dimensional surface to the residuals of
\Teff, $\log g$, and Fe/H between the calibrated Keck and initial APF
parameters. The errors on the stellar parameters are a quadrature sum
of the statistical errors from the DE-MCMC fits and the scatter in the
APF vs.\ Keck calibration. The scatter in the calibration is generally
an order of magnitude larger than the statistical errors in the S/N
regime for the \ktwo targets observed on APF.

\cite{petigura:2015phd} assessed the accuracy of
\texttt{SpecMatch}-derived stellar parameters by modeling the spectra
of several samples of touchstone stars with well-measured
properties. The properties of these stars were determined from
asteroseismology \citep{huber:2012}, detailed LTE spectral modeling
and transit light curve modeling \citep{torres:2012}, and detailed LTE
spectral modeling \citep{valenti:2005}. The uncertainties of \specmatch
parameters are dominated by errors in the \cite{coelho:2005} model
spectra (e.g., inaccuracies in the line lists, assumption of LTE,
etc.). Given that we observe spectra at S/N $\gtrsim 35$ per pixel,
photon-limited errors are not an appreciable fraction of the overall
error budget.

To estimate stellar masses and radii for all stars with
\texttt{SpecMatch} parameters, we use the free and open source
\texttt{isochrones} Python package \citep{morton:2015b}.  This tool
accepts as inputs the \Teff, \ensuremath{\log g}, and
\ensuremath{[\mbox{Fe}/\mbox{H}]} measured by \texttt{SpecMatch} and
interpolates over a grid of stellar models from the Dartmouth Stellar
Evolution Database \citep{dotter:2008}.  \texttt{isochrones} uses the
\texttt{emcee} Markov Chain Monte Carlo package
\citep{foreman-mackey:2012} to estimate uncertainties, sometimes
reporting fractional uncertainties as low as 1\%.  Following
\citet{sinukoff:2016}, we adopt a lower limit of 5\% for the
uncertainties on stellar mass and radius to account for the intrinsic
uncertainties of the Dartmouth models found by \citet{feiden:2012}.

85 stars in our sample lack \texttt{SpecMatch} parameters. For most of
these, we adopt the stellar parameters of \cite{huber:2016}.  This
latter analysis relies on the Padova set of stellar models
\citep{marigo:2008}, which systematically underestimate the stellar
radii of low-mass stars.  Follow-up spectroscopy to provide refined
parameters for these later-type stars is underway (Dressing et al., in
prep.; Martinez et al., in prep.).  Our sample includes a small number
of stars not considered by \cite{huber:2016}, such as targets in
\ktwo's Campaign 0. For these, we use \texttt{isochrones} in
conjunction with broadband photometry collected from the APASS, 2MASS,
and WISE surveys to infer the stellar parameters.



We then use the free, open-source \texttt{LDTk} toolkit
\citep{parviainen:2015} to propagate our measured \Teff, $\log g$,
      [Fe/H], and their uncertainties into limb-darkening coefficients
      and associated uncertainties. These limb-darkening parameters
      act as priors in our transit light curve analysis (described
      below in Sec.~\ref{sec:tlc}). We upgraded \texttt{LDTk} to allow
      the (typically non-Gaussian) posterior distributions generated
      by the \texttt{isochrones} package to be fed directly into the
      limb darkening analysis\footnote{GitHub commits 60174cc,
        46d140b, and 8927bc6}.  Because \texttt{LDTk} often reports
      implausibly small uncertainties on the limb darkening
      parameters, based on our experience with such analyses we increase all these uncertainties by a factor of
      five in our light curve analyses. Spot checks of a number of
      systems reveal that imposing priors on the stellar
      limb-darkening has a negligible impact ($<1\sigma$) on our final
      results, relative to analyses with much weaker constraints on
      limb darkening.

All our derived stellar parameters --- \Teff, \logg, \Rstar,
\Mstar --- and their uncertainties are listed in Table~\ref{tab:stars}.

\section{Transit Light Curve Analyses}
\label{sec:tlc}
After identifying planet candidates and determining the parameters of
their host stars, we subject the detrended light curves to a full
maximum-likelihood and MCMC analysis. We use a custom Python wrapper
of the free, open source \texttt{BATMAN} light curve code
\citep{kreidberg:2015b}. We upgraded the \texttt{BATMAN} codebase to
substantially increase its efficiency when analyzing long-cadence
data\footnote{GitHub commit 9ae9c83}. The light curves are fit using
the standard Nelder-Mead Simplex Algorithm\footnote{As implemented in
  \texttt{scipy.optimize.fmin}} and then run through \texttt{emcee} to
determine parameter uncertainties.

Our general approach follows that used in our previous papers
\citep{crossfield:2015a,petigura:2015a,schlieder:2016,sinukoff:2016}. The
model parameters in our analysis are the transit time ($T_0$); the
candidate's orbital period and inclination ($P$ and $i$); the scaled
semimajor axis ($a/R_*$); the fractional candidate size ($R_p/R_*$);
the orbital eccentricity and longitude of periastron ($e$ and
$\omega$), the fractional level of dilution ($\delta$) from any other
sources in the aperture; a single multiplicative offset for the
absolute flux level; and quadratic limb darkening coefficients ($u_1$
and $u_2$).  We initialize each fit with the best-fit parameters
returned from \TERRA.  Note that both this analysis and that of \TERRA
assume a linear ephemeris, so systems with large TTVs could be missed
or misidentified.

During the analysis, several parameters are constrained or subjected
to various priors. Gaussian priors are applied to the limb-darkening
parameters (as derived from the \texttt{LDTk} analysis), to $P$ (with
a dispersion of $\sigma_P=0.01$~d, to ensure that the desired
candidate signal is the one being analyzed), and to $e$
($\mu_e=10^{-4}$ and $\sigma_e=10^{-3}$, to enforce a
circular orbit). We also apply a uniform prior to $T_0$ (with width
$0.06P$), $i$ (from 50\degree\ to 90\degree), $R_p/R_*$ (from $-1$ to
1), and $\omega$ (from 0 to 2$\pi$); both $P$ and $a/R_*$ are
furthermore constrained to be positive. Allowing $R_P/R_*$ to take on
negative values avoids the Malmquist bias that would otherwise result
from treating it as a positive-definite quantity. For those systems
with no identified stellar companions, our high-resolution imaging
and/or spectroscopy constrain the dilution level; otherwise, we adopt
a log-uniform prior on the interval ($10^{-6}$, 1).

\section{False Positive Assessment}
\label{sec:fpp}

During the prime \Kepler mission, both the sheer number of planet
candidates and their intrinsic faintness made direct confirmation by
radial velocities impractical for most systems.  Nonetheless many
planets can be statistically validated by assessing the relative
probabilities of planetary and false positive scenarios; a growing
number of groups have presented frameworks for quantitatively
assessing the likelihoods of planetary and false positive scenarios
\citep{torres:2011,morton:2012,diaz:2014,santerne:2015}. These false
positive scenarios come in several classes: (1) undiluted
eclipsing binaries, (2) background (and foreground) eclipsing binaries
where the eclipses are diluted by a third star, and (3) eclipsing
binaries in gravitationally-bound triple systems. 

To estimate the likelihood that each of our planet candidates is a
true planetary system or a false positive configuration we use the
free and open source \texttt{vespa} software
\citep{morton:2015}. \texttt{vespa} compares the likelihood of each
scenario against the planetary interpretation and accepts additional
constraints from HRI and spectroscopy.  Throughout this analysis, we
apply Version 0.4.7 of \texttt{vespa} (using the MultiNest backend) to
each individual planetary candidate.  Other types of false positive
scenarios exist that are not explicitly treated by \texttt{vespa}, such
extremely inconvenient
arrangements of starspots.  The community's experience of following up
transiting planet candidates indicates that such scenarios are much
less common than the dominant arrangements considered by \texttt{vespa}; nonetheless
quantifying the likelihood of such scenarios for each candidate would
be an interesting avenue for future research.

\subsection{Calculating FPPs}
To calculate the False Positive Probability (FPP) for each system, we
use as inputs: stellar photometry from APASS, 2MASS, and WISE; the
stellar parameters described in Sec.~\ref{sec:stars}; the detrended
light curve (after masking out any transits from other candidates in
that system); the exclusion constraints from adaptive optics imaging
data in terms of contrast vs.\ separation (where available) and from
our high-resolution spectroscopy (maximum allowed contrast and
velocity offset); and an upper limit on the depth of any secondary
eclipse.  We derive the last of these by constructing a rectangular
signal with depth unity and duration equal to the best-fit transit
duration, scanning the template signal across the out-of-transit light
curve, and reporting the 99.7$^{th}$ percentile as the eclipse depth's
upper limit.

We report the final False Positive Probabilities (FPPs) of all our
systems in Table~\ref{tab:results}.  For the purposes of the
discussion that follows, we deem any candidate signal with
FPP$<$\minfpp as a  validated extrasolar planet and
signals with FPP$>$\maxfpp as false positives. For all unvalidated
candidates, Table~\ref{tab:fppbreakdown} summarizes \texttt{vespa}'s
estimate of the relative (unnormalized) likelihood of each potential
false positive scenario.


The \texttt{vespa} algorithm implicitly assumes that each planet
candidate lacks any other companion candidates in the same
system. Studies of \Kepler's multiple-candidate systems show that
almost all are planets \citep{lissauer:2012}.  This ``multiplicity
boost'' has subsequently been used to validate hundreds of
multi-planet systems \citep{rowe:2014}.  Because \texttt{vespa} treats
only single-planet systems, we simply treat these multi-candidate
systems as independent, isolated candidates in the FPP
analysis. \cite{sinukoff:2016} show that \ktwo's multiplicity boost is
$\ge20$ even in crowded fields, comparable to the boost factor derived
for the original \Kepler mission.

Even without the multiplicity boost, our approach validates the
majority of our multi-candidate systems.  Both EPIC~201445392 (K2-8)
and EPIC~206101302 host two planet candidates. In each system we
validate one candidate and find FPP\,=\,4--7\%\ for the
other. The \ktwo multiplicity boost factor of $\ge20$ therefore
results in all candidates in both systems being firmly labeled as
validated planets.

A more complicated case is EPIC~205703094, which hosts three planet
candidates. Our \texttt{vespa} analysis finds that one candidate is a
false positive and that the others both have FPP\,$\approx$\,50\% (see
Tables~\ref{tab:results} and~\ref{tab:fppbreakdown}). Our light curve
analysis finds all three candidates are well-fit by grazing transits
($b\sim1$), leaving $R_P/R_*$ only weakly constrained. Furthermore,
our high-resolution imaging reveals that the system is a close visual
binary with separation 0.14'' (see Tables~\ref{tab:aoval}
and~\ref{tab:aocomp}).  Therefore we can neither validate nor rule out
the three candidates in this system.

\subsection{Targets with nearby stellar companions}
\label{sec:dilution}
Planet candidates orbiting stars in physical or visual multiple
systems are much more difficult to validate due to blending in the
photometric aperture \citep[see
e.g.][]{ciardi:2015}. Table~\ref{tab:aocomp} shows that our \ktwo
photometric apertures are quite large (up to 20'' in extreme cases)
and that HRI follow-up reveals stellar companions within these
apertures for many systems. Therefore we must treat these systems with
greater care.

To demonstrate the difficulty, consider  two stars with flux ratio
$F_2/F_1<1$ and angular separation $\rho$. Assume both lie in a
photometric aperture with radius $r>\rho$, with which a transit is
observed with apparent depth $\delta'$. If the transit occurs around
the primary star, then the true transit depth is $\delta_1 \approx
\delta' / \left( 1 - F_2/F_1 \right)$; this is at most twice the
observed depth, indicating a planetary radius up to $\sqrt{2}$ larger
than otherwise determined.  If instead the transiting object orbits
the secondary, then the true transit depth is $\delta_2 \approx
\delta' F_1/F_2$ and the transiting object may be many times larger
than expected.  Table~\ref{tab:aoval} lists all candidates known to
host secondary stars and their relationships between $F_2/F_1$ \&
$\delta'$ and $\rho$ \& $r$.

Any planet candidate in a multi-star system and with $F_2/F_1<\delta'$
cannot transit the secondary (which is too faint to be the source of
the observed transit signal). We find several such systems, though
only two
(EPIC~202126852 and 211147528) have FPP$<$0.95. Nonetheless, for
all these systems we account for the dilution of the secondary star(s)
as described below. 

For candidates with $\delta' < F_2/F_1$ and $\rho < r$, the transit
could occur around either star. We compare our nominal time series
photometry to that computed with $r=1$\,pixel for all such
candidates. For targets with more widely-separated nearby stars, if
the one-pixel-photometry reveals a shallower transit then the transit
probably occurs around the secondary star. However, if $\rho<1$\,pix
then we cannot reliably identify the source of the transits. We find
\nunval candidates of these types that we cannot validate at present, and
note the disposition of all such systems in Table~\ref{tab:aoval}.

For all remaining systems, the detected transits must occur around the
primary star but will be diluted by light from the secondary.  We
estimate the total brightness of these systems' secondary star(s) as
follows. For stars detected by optical imaging (Robo-AO and DSSI), we
use the measured contrast ratio with an uncertainty of 0.05~mag.  For
stars detected by infrared imaging, we use the relations of
\cite{howell:2012} to translate the observed infrared color into the
\Kepler bandpass. Since these relations are approximate and depend
strongly on spectral type, we conservatively apply an uncertainty of
0.5~mag to these values. Sec.~\ref{sec:dilution} describes how we use
these data to constrain the dilution parameter's posterior
distribution, thereby reducing the systematic biases induced by
unrecognized sources of dilution \citep[e.g.,][]{ciardi:2015}.



\section{Results and Discussion }
\label{sec:discussion}

We find \nplanet validated planets (i.e., FPP\,$<$\,\minfpp) in our set of
\ntotcand planet candidates.  Significantly, we show that \ktwo's
surveys increase by $30\%$ the number of small planets orbiting
moderately bright stars compared to previously known planets.  Below
in Sec.~\ref{sec:overview} we present a general overview of our survey
results. Then in Sec.~\ref{sec:individual} we discuss individual
systems, both new targets and previously identified planets and
candidates.







%

%

\subsection{Overview of Results}
\label{sec:overview}
\begin{figure}[ht!]
\begin{center}
\includegraphics[width=3.5in]{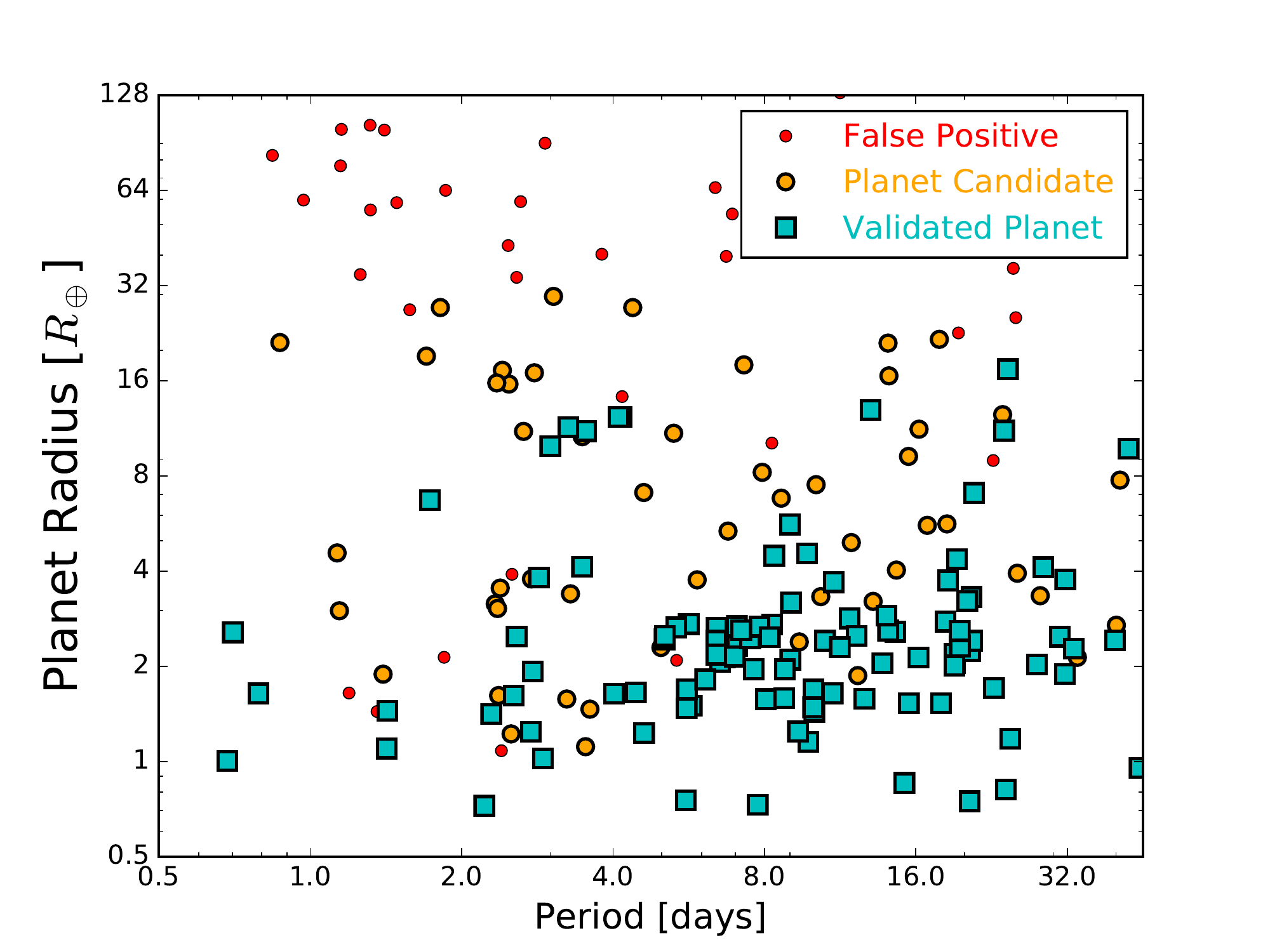}
\caption{\label{fig:period_radius} Orbital periods and radii of our
  \nplanet validated planets, \nfp false positive systems, and
  \nfinalcand remaining planet candidates. Uncertainties on planet
  radius (listed in Table~\ref{tab:results}) are typically
  $\sim$13\%.  }
\end{center}
\end{figure}

Our validated planetary systems span a range of properties, with
median values of $R_P$=\medrad, $P$=\medper, \Teff\,=\,\medteff, and
Kp=\medkp. Fig.~\ref{fig:period_radius} shows the distribution of
planet radius, orbital period, and final disposition for our entire
candidate sample.  The candidates range from 0.7--44\,d,
and from $<1R_\oplus$ to larger than any known planets. 

\begin{figure}[ht!]
\begin{center}
\includegraphics[width=3.5in]{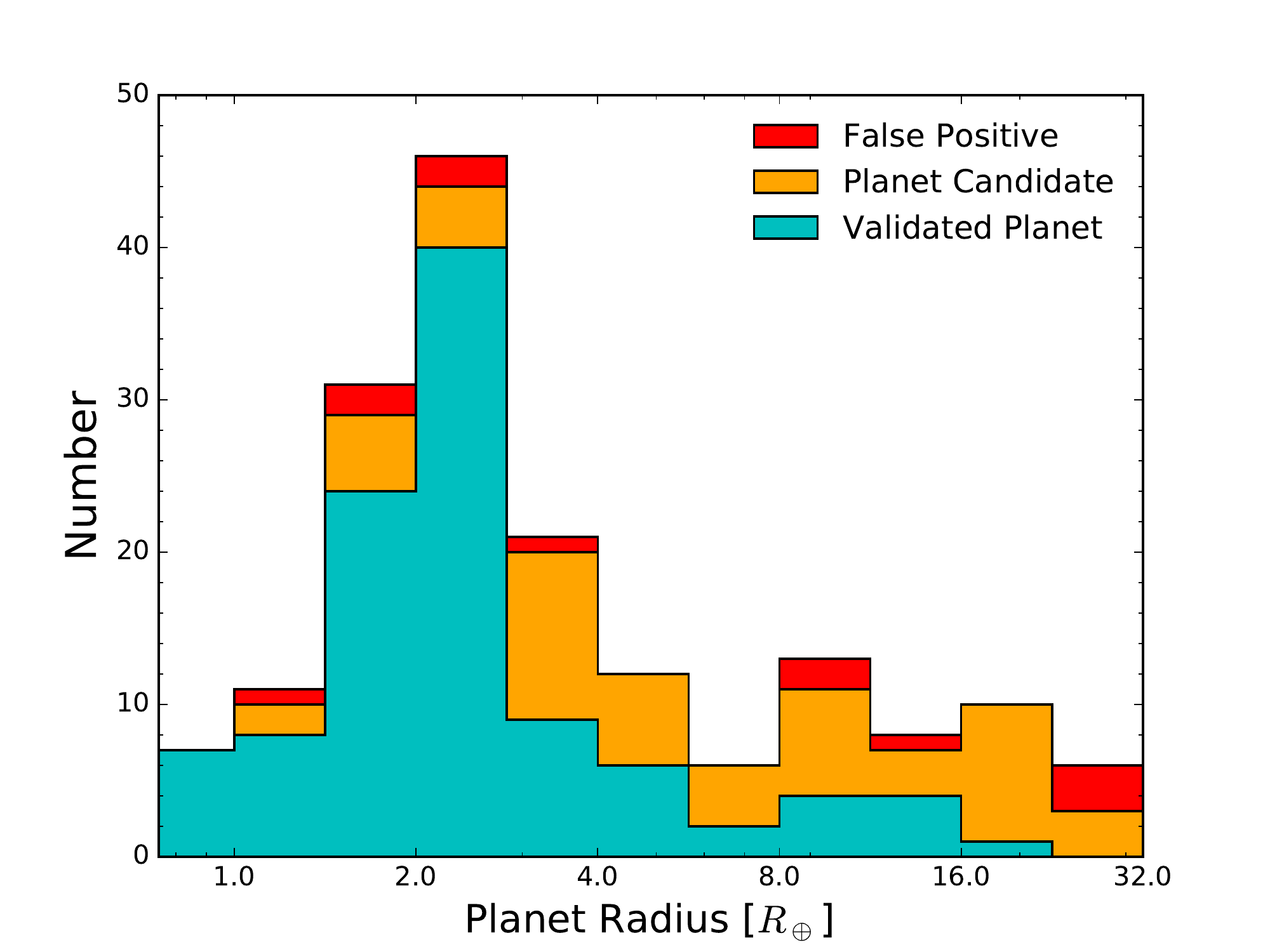}
\caption{\label{fig:radius_dist} Distribution of  planet candidate
  radii for our validated planets, false positive systems, and
  remaining planet candidates.  We validate most of the candidates
  smaller than $3\,R_\oplus$, consistent with the  low false positive
  rate we find for small planets.}
\end{center}
\end{figure}

\begin{figure}[ht!]
\begin{center}
\includegraphics[width=3.5in]{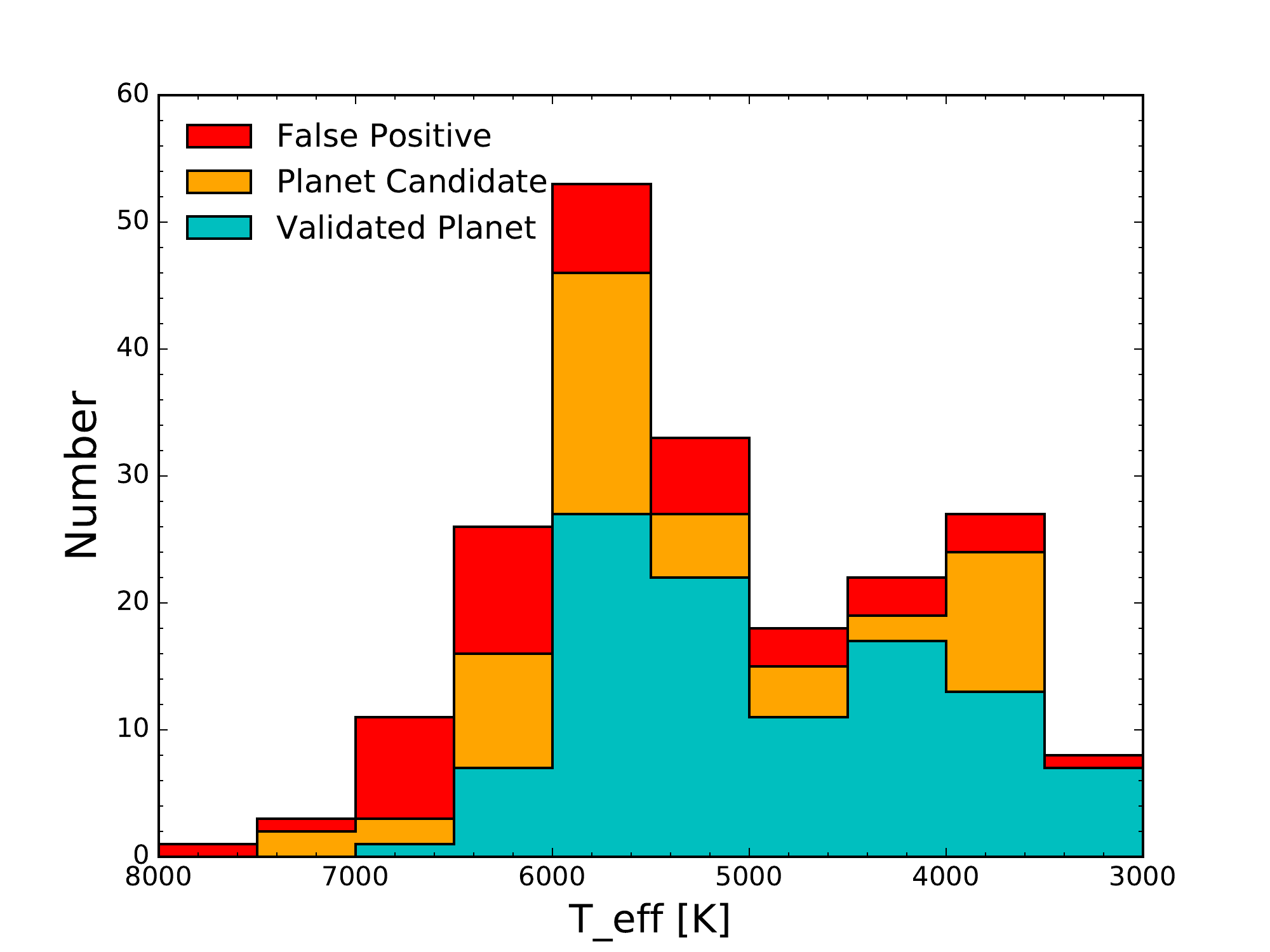}
\caption{\label{fig:teff_dist} Distribution of stellar
  effective temperatures for systems with validated planets, false
  positive, and remaining planet candidates.  There is a hint of a
  higher validation rate around stars cooler than $\sim5500$\,K.}
\end{center}
\end{figure}

\begin{figure}[ht!]
\begin{center}
\includegraphics[width=3.5in]{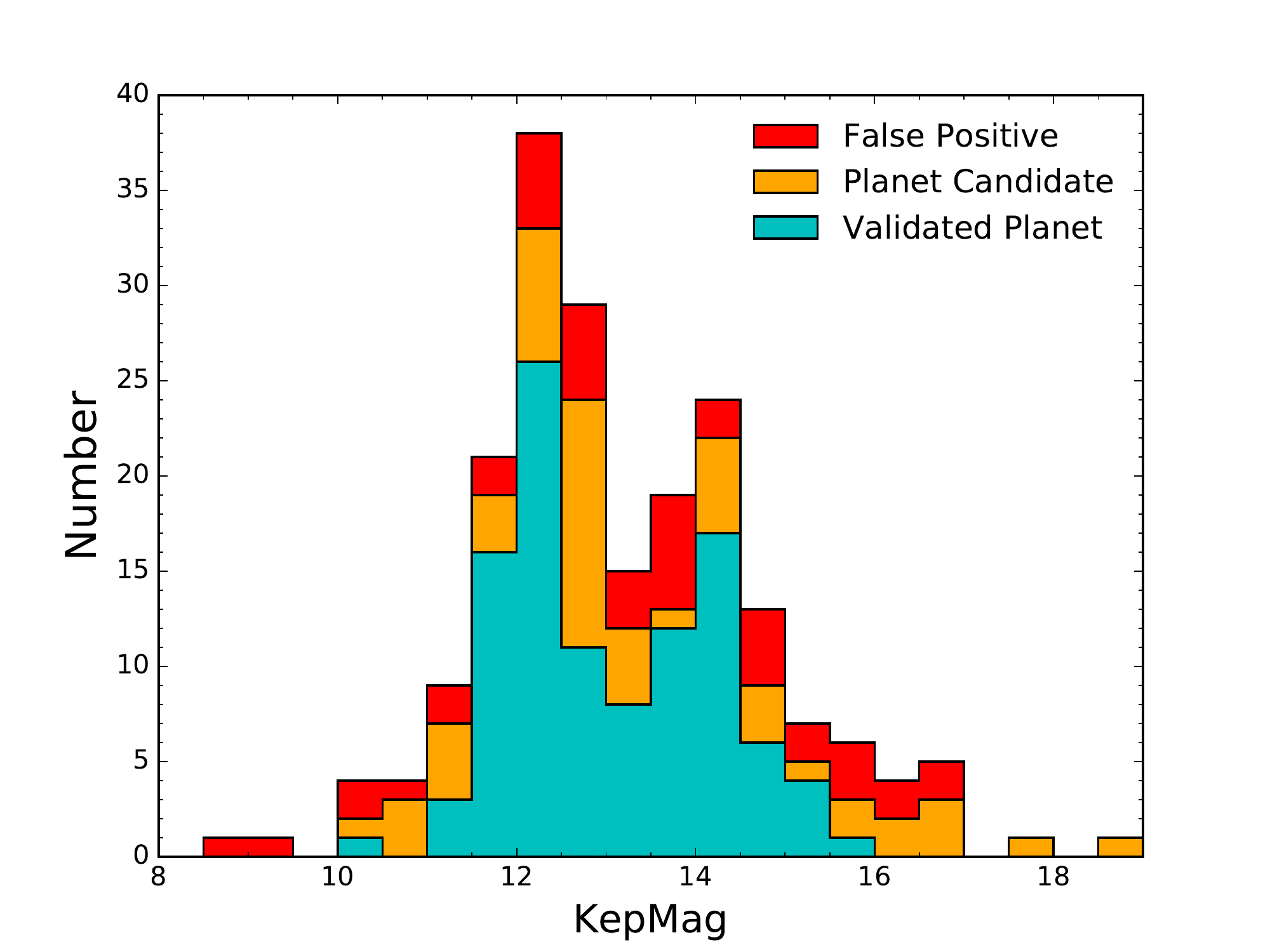}
\caption{\label{fig:kepmag_dist} Distribution of {\em Kp} for
  systems with validated planets, false positives, and remaining
  planet candidates.  Our brightest validated system, K2-65 (EPIC 206144956),
  contains a 1.6\,\rearth planet orbiting a V=10.8~mag star.}
\end{center}
\end{figure}

Fig.~\ref{fig:radius_dist} shows that the majority of candidates have
$R_P<3$\rearth, and these smallest candidates exhibit the highest
validation rates.  In contrast, we validate less than half of
candidates with $R_P>3$\rearth and less than half of candidates with
$P<2$\,d (Fig.~\ref{fig:period_dist}).  We find a substantially higher
validation rate for target stars cooler than $\sim5500$\,K vs.\ hotter
stars (65\% vs.\ 37\%; see Fig.~\ref{fig:teff_dist}).
Fig.~\ref{fig:kepmag_dist} shows that we validate no systems with {\em
  Kp}\,$>16$\,mag, but otherwise reveals no obvious trends with
stellar brightness.

Our analyses leave \nfinalcand planet
candidates with no obvious disposition (i.e.,
\minfpp$<$\,FPP\,$<$\,\maxfpp).  These candidates are typically large
($R_P>3\Re$), and their FPPs are listed in Table~\ref{tab:results}.
Furthermore, in Table~\ref{tab:fppbreakdown} we list the individual
likelihoods of each false positive scenario considered by
\texttt{vespa}.  

We calculate the false positive rate (FPR) of our entire planet
candidate sample by taking our \ntotcand candidates, excluding the
\nunval candidates with nearby stars discovered by HRI that we cannot
validate (see Sec.~\ref{sec:dilution}), and integrating over the
probability that each candidate is a planet. In this way we estimate
that our entire sample contains roughly 145 total planets (though we
validate just \nplanet). This ratio corresponds to a false positive
rate of  15--30\%, with higher FPPs for
candidates showing larger sizes and/or shorter orbital periods (see
Figs.~\ref{fig:period_dist} and~\ref{fig:radius_dist}).

We also split our sample into several bins in radius and period to
estimate the FPR for each subset, listed in Table~\ref{tab:fprates}.
Our false positive rate is dominated by larger candidates, just as
Fig.~\ref{fig:radius_dist} suggests. Sub-Jovian candidates (with $R_P\le
8R_\oplus$) have a cumulative FPR of $\sim 10\% $, whereas over half
of larger candidates are likely false positives. The FPR for larger
candidates is consistent with that measured for the original \Kepler
candidate sample \citep{santerne:2016}.  Candidates with $P<3$\,d have
a FPR roughly twice as high as for longer-period systems.

Since we have excluded the \nunval candidates described above, these
FPRs are only approximate and we defer a more detailed analysis of our
survey completeness and accuracy to a future publication.
Nonetheless, further follow-up observations for systems lacking
high-resolution spectroscopy, high-resolution imaging, and/or RV
measurements may expect to identify, validate, and confirm a
considerable number of additional planetary systems.

\begin{figure}[ht!]
\begin{center}
\includegraphics[width=3.5in]{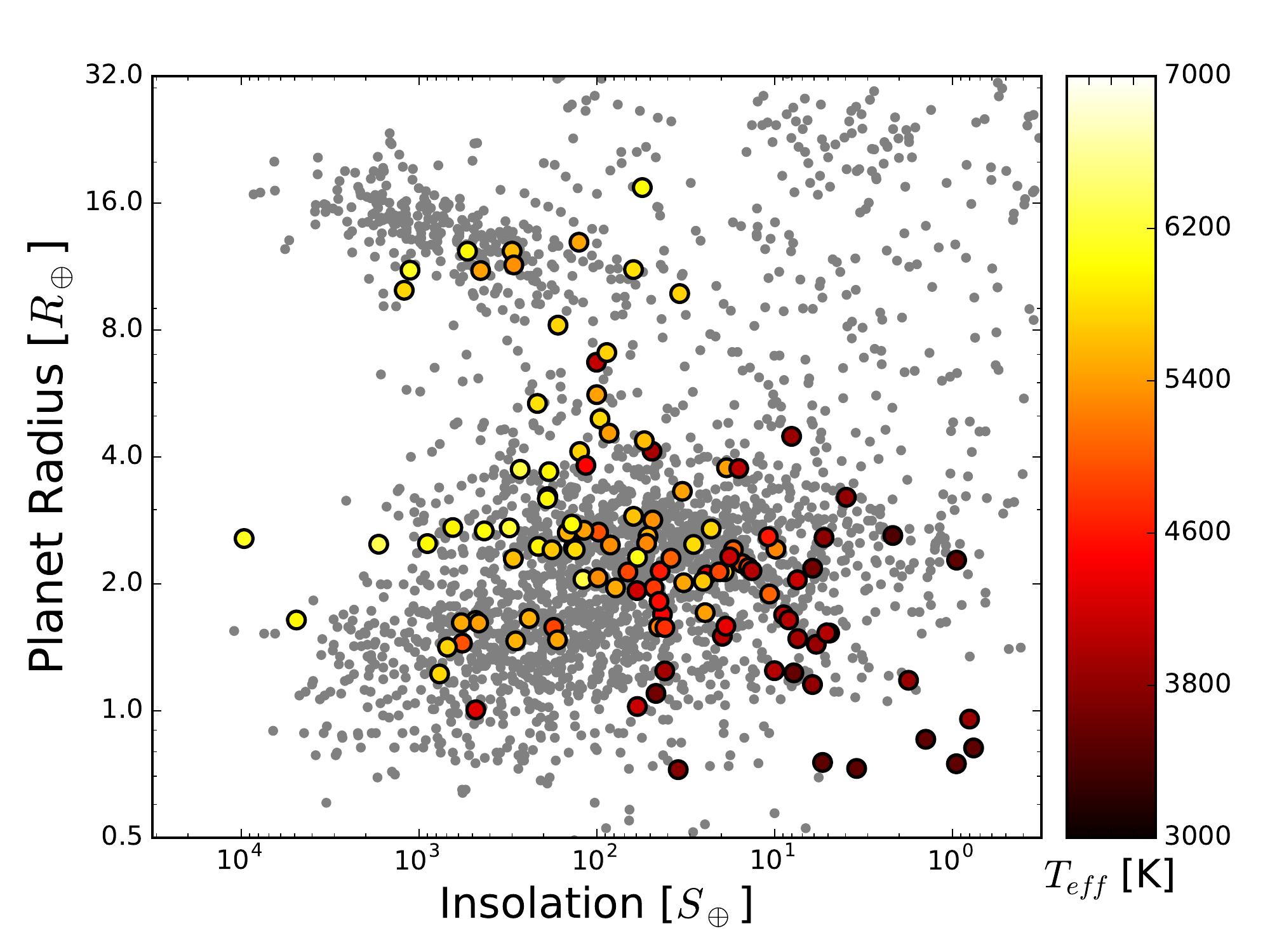}
\caption{\label{fig:insolation_radius_teff} Planetary radii, incident
  insolation, and stellar effective temperature for our \nplanet
  validated planets (colored points) and all planets at the NASA
  Exoplanet Archive (gray points).  As expected, most of our smaller,
  cooler planets are found around cooler, later-type stars ($\Teff
  \lesssim 4000$\,K).  Uncertainties, omitted for clarity, are listed
  in Table~\ref{tab:results}.  Statistical uncertainties on planet
  radius and insolation (listed in Table~\ref{tab:results}) are
  typically $\sim$13\% and $\sim$26\%, respectively, but the coolest
  host stars are likely larger, hotter, and more luminous than they
  appear \citep{huber:2016}.  }
\end{center}
\end{figure}

Fig.~\ref{fig:insolation_radius_teff} shows planet radius versus the
irradiation levels incident upon each of our validated planets
relative to that received by the Earth ($S_\oplus$), color-coded by
\Teff.  These planets receive a wide range of irradiation, from
roughly that of Earth to over $10^4\times$ greater. As
expected, our coolest validated planets orbit cooler stars (K and M
dwarfs).  However, we caution that the stellar parameters for these
systems come from broadband colors and/or \cite{huber:2016}, so
uncertainties are large and biases may remain. Follow-up spectroscopy
is underway to more tightly constrain the stellar and planetary
properties of these systems (Dressing et al., in prep; Martinez et
al., in prep.).

Finally, Fig.~\ref{fig:jmag_radius_teff} shows that \ktwo planet
survey efforts have substantially increased the number of smaller planets
known to orbit moderately bright stars.  Although our sensitivity appears to drop off
below $\sim1.3$\rearth (as shown in Fig.~\ref{fig:radius_dist}) and we
find no planets around stars brighter than $J<8.9$~mag, we validate a
substantial number of intermediate-size planets around moderately
bright stars. In particular, the right-hand panel of
Fig.~\ref{fig:jmag_radius_teff} shows that the first five fields of
\ktwo have already increased the number of small planets orbiting
fairly bright stars by roughly 30\% compared to those tabulated at the
NASA Exoplanet Archive. Considering the sizes of these planets and
  the brightness of their host stars, many of these systems are
  amenable to follow-up characterization via Doppler spectroscopy
  and/or JWST transit observations.

\input{fprate_results_v6.tex}



\subsection{Notes on Individual Systems}
\label{sec:individual}

Of the \nplanet planets validated by our analysis, \newplanet are
newly validated.  These include several new multi-planet systems,
systems as bright as V=10.8~mag, and several small, roughly
Earth-sized planets receiving roughly Earth-like levels of
irradiation. Below we describe some of the most interesting new
systems in Sec.~\ref{sec:newval}, our analysis of previously confirmed
or validated planets in Sec.~\ref{sec:prevconf}, and our results for
known but unvalidated candidates in Sec.~\ref{sec:prev}.

\subsubsection{New Validated Planets}
\label{sec:newval}

\begin{figure}[ht!]
\begin{center}
\includegraphics[width=5.5in]{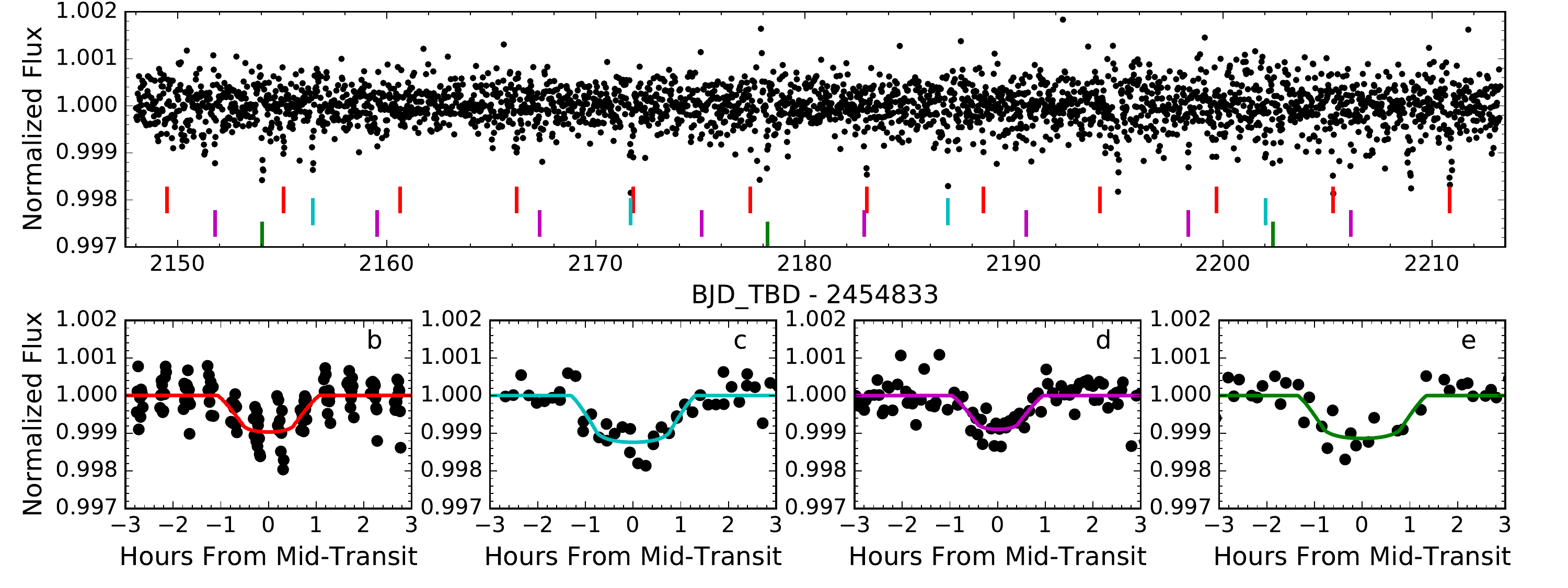}
\caption{\label{fig:multi_lightcurve} Photometry of K2-72 (EPIC~206209135),
  which hosts four transiting planets. {\em Top:} Full time series
  with colored tick marks indicating each individual transit time.
  {\em Bottom:} Phase-folded photometry with the color-coded, best-fit
  transit model overplotted for each planet. Our analysis indicates a
  stellar radius of $0.40^{+0.12}_{-0.07}$\,\rsun, planetary radii of
  1.2--1.5\,\rearth, and (from left to right) orbital periods of 5.58,
  7.76, 15.19, and 24.16\,d. }
\end{center}
\end{figure}

K2-72 (EPIC~206209135) is a dwarf star hosting a planet candidate on a 5.57\,d
orbit \citep{vanderburg:2016}; we find three additional candidates and
validate all four planets in this system. We see the transits in both
our photometry (shown in Fig.~\ref{fig:multi_lightcurve}) and that of
\cite{vanderburg:2014}, and our light curve fits give consistent
values of $\rho_{*,circ}$ for all planets -- both points give us
confidence that these are true planetary systems. \cite{huber:2016}
reports a stellar radius of 0.23\,\rsun\ but notes that this is likely
an underestimate. The weighted mean of our four stellar density
measurements is $9.0\pm3.6$~g~cm$^{-3}$; using the mass-radius
relation of \cite{maldonado:2015} implies a stellar radius of
$0.40^{+0.12}_{-0.07}$\,\rsun and planetary radii of
1.2--1.5\,\rearth\ for all planets. Analysis of the stellar spectrum
is also consistent with this size (Martinez et al., in prep.; Dressing
et al., in prep.). These four small planets have orbital periods of
5.58, 7.76, 15.19, and 24.16\,d.  The irradiation levels for several
planets are also quite consistent with Earth's insolation.  Several of
these planet pairs orbit near mean motion resonances: planets c and d
orbit near the first-order 2:1 MMR, and b and c orbit near the
second-order 7:5 MMR.  Although the star K2-72 is relatively faint ---
$Kp=14.4$~mag, $K=11.0$~mag --- and so follow-up Doppler or transit spectroscopy observations
to measure the planets' masses or atmospheric compositions will be
challenging, the system's near-integer period ratios suggest that
measurements of transit timing variations (TTV) may help reveal the
masses and bulk densities of all these planets.

We also identify and validate four new two-planet systems in Campaign
4: K2-80, K2-83, K2-84, and K2-90 (EPIC 210403955, 210508766,
210577548, and 210968143, respectively).  Our light curve analyses of
the planets in each system yield values of $\rho_{*,circ}$ that are
consistent at $<1\sigma$, consistent with the hypothesis that both
planets in each pair orbit the same star. Future transit followup of
these systems will be challenging but feasible, with the most easily
observed transits having depths of $\sim$1\,mmag. None of the systems
appear to have planets near low-order mean-motion resonance, but
additional (non-transiting) planets in these systems could lie near
resonance and induce detectable TTVs.

Our brightest validated system, K2-65 (EPIC 206144956), contains a
1.6\,\rearth planet orbiting a star with V=10.8~mag, J=9.0~mag located
in Campaign 3. Despite its 13\,d orbital period and low predicted
radial velocity semiamplitude (likely $\lesssim1$m~s$^{-1}$), the
bright star, relatively small planet size, and low planet insolation
(just $45\times$ that of Earth's) may make this system an attractive
target for future RV efforts.

Also of interest for radial velocity followup is K2-89 (EPIC
210838726), which hosts a highly irradiated, roughly Earth-sized
planet on a one-day orbit around an M dwarf.  The planet should have a
radial velocity semi-amplitude of roughly 1~m~s$^{-1}$, and although
the star is not especially bright (Kp=13.3~mag, K=10.1~mag) detection
of the planet's RV signal may lie within reach of existing and planned
high-precision Doppler spectrographs.

\subsubsection{Previously Confirmed Planets}
\label{sec:prevconf}


K2-3bcd and K2-26b (EPIC~201367065 and 202083828, respectively) were
previously validated as sub-Neptune-sized planets orbiting M dwarfs
\citep{crossfield:2015a,schlieder:2016}, and K2-3b was confirmed by
Doppler spectroscopy \citep{alemenara:2015}. Transits of all four
planets were also recently observed by {\em Spitzer}
\citep{beichman:2016}. We independently validate all these planets. Note however
that the stellar parameters we estimate here for these systems
systematically underestimate the more accurate,
spectroscopically-derived parameters presented in those papers.

K2-10b and K2-27b (EPIC 201577035b and 201546283b, respectively) were
previously validated as planets \citep{montet:2015,vaneylen:2016}. We
find FPP\,$<$\,\minfpp for both, thus independently validating these two
planetary systems.  A new stellar companion with $\rho=3$'' and
$\Delta i=5.8$~mag slightly dilutes the latter's transit but does not
significantly affect its reported parameters.

We report a new stellar companion with $\rho=3.2$'' and $\Delta
K=5.8$~mag near K2-13b \citep[EPIC 201629650;][]{montet:2015}.  This
new, faint star is bright enough that it could be the source of the
observed transits; we therefore suggest that this previously-validated
system should be deemed a planet candidate.

WASP-47 (EPIC 206103150) hosts a hot Jupiter \citep[planet
  b;][]{hellier:2012}, a giant planet on a 1.5\,yr orbit
\citep[c;][]{neveu-vanmalle:2016}, and two additional, short-period
transiting planets \citep[d and e;][]{becker:2015}.  Our analysis of
the three transiting planets yields FPP\,$<$\,\minfpp for each, so we
independently validate this planetary system.  

HAT-P-56b (EPIC 202126852b) is a hot Jupiter confirmed by measuring
the planet's mass with Doppler spectroscopy \citep{huang:2015}.  Our
analysis indicates that the planetary hypothesis is the most probable
explanation for the signal detected, with the next-most-likely
scenario being an eclipsing binary (FPP=65\%; see
Table~\ref{tab:fppbreakdown}).  However, the radial velocity
measurements of \cite{huang:2015} rule out the eclipsing binary
scenario favored by \texttt{vespa} and so confirm the planetary nature
of this system.

K2-19b and~c (EPIC~201505350bc) were identified as a pair of planets
with an orbital period commensurability near 3:2 \citep[7.9\,d and
  11.9\,d;][]{armstrong:2015,narita:2015,barros:2015}.  A third
candidate with period 2.5\,d was subsequently identified and validated
\citep{sinukoff:2016}, which is not near any low-integer period ratios
with the previously identified planets. Our analysis independently validates all
three of these planets.

K2-21b and~c (EPIC-206011691bc) are two planets with radii 1.5--2\,\Re
orbiting near a 5:3 orbital period commensurability
\citep{petigura:2015a}, and K2-24 b and~c (EPIC 203771098bc) are two
low-density sub-Saturns orbiting near a 2:1 orbital period
commensurability and with masses measured by Doppler spectroscopy
\citep{petigura:2016a}.  Our analysis yields FPP\,$<$\,\minfpp for all
four of these planets, thereby confirming their planetary status.

K2-22b (EPIC 201637175b) is a short-period rocky planet caught in the
act of disintegrating in a 9\,hr period around its host star
\citep{sanchis-ojeda:2015}. Our analysis successfully identifies this
as a planet candidate and \texttt{vespa} indicates that the planetary
hypothesis is the most probable explanation for the signal detected
(FPP=15\%; see Table~\ref{tab:fppbreakdown}). However, because
\texttt{vespa} cannot account for this system's highly variable
transit depths (from 1\% to as shallow as $<10^{-3}$) the measured FPP
is not reliable. We do not claim to de-validate K2-22b.

K2-25b (EPIC~210490365) is a Neptune-size planet transiting an M4.5
dwarf in the Hyades \citep{mann:2016,david:2016b}. In our transit
search, \TERRA locked on to this star's 1.8\,d rotation period and so
we did not identify the planet candidate.

K2-31b (EPIC~204129699b) is a hot Jupiter validated by radial velocity
spectroscopy \citep{grziwa:2016,dai:2016}. Because of the grazing
transit the planet radius is only poorly determined.  The best-fit
planet radius listed in Table~\ref{tab:results} is implausibly large
given the measured mass; this large radius likely led the
\texttt{vespa} analysis to incorrectly assign this confirmed planet a
FPP of 84\%.

EPIC 206318379b was validated by \cite{hirano:2016} as a
sub-Neptune-sized planet transiting an M dwarf. We did not identify
the system in our transit search; a subsequent investigation shows
that our photometry and transit search code did not properly execute for this
system, and was never restarted.

K2-29b and K2-30b (EPIC~211089792 and 210957318) are hot Jupiters
whose masses were recently measured via Doppler spectroscopy
\citep{johnson:2016,lillo-box:2016,santerne:2016b}. We find FPP\,$<$\,\minfpp for
both, and so independently validate these systems.

The Sun-like star BD+20 594 (EPIC~210848071) is reported to host a planet with
radius 2.3\,$R_\oplus$ on a 42\,d orbit \citep{espinoza:2016}. Since
\ktwo observed only two transits of this planet, our transit search
did not identify this system (see Sec.~\ref{sec:vetting}).

The first large sets of planet candidates and validated planets from
\ktwo were produced by \cite{foremanmackey:2015} and
\cite{montet:2015}.  The former identified 36 planet candidates, of
which the latter validated 21.  We successfully independently validate all but
two of these planets, and find that for both outliers the
disagreements are marginal. For K2-8b (EPIC~201445392b) we measure
FPP\,=\,4.2\%, but as discussed in Sec.~\ref{sec:fpp} the multiplicity
boost  suppresses this candidate's  FPP and  results in a validated planet.
However, we measure FPP\,=\,45\% for K2-9b (EPIC~201465501), almost
10 times greater than originally reported. We attribute this
difference to the stellar parameters reported for this star from our
homogeneous \texttt{isochrones} stellar analysis: it reports K2-9 to
be an early M dwarf, but a more reliable spectroscopic analysis
reveals the star to be a smaller and cooler mid-M dwarf
\citep{schlieder:2016,huber:2016}. The planet K2-9b is successfully
validated when we use the spectroscopic parameters in our FPP analysis
\citep{schlieder:2016}. The discrepancy highlights the importance of
accurate and spectroscopically-derived stellar parameters (especially
for M dwarfs) when assessing planetary candidates.

Recently, our team validated several new multi-planet systems found by
\ktwo: K2-35, -36, -37, and -38 \citep[EPIC 201549860, 201713348,
  203826436, and 204221263;][]{sinukoff:2016}. Our analysis here uses
much of the same machinery as in that work, so it should be little
surprise that we again validate all planets in these systems.

Most recently, the giant planet K2-39b was confirmed by radial
velocity spectroscopy \citep{vaneylen:2016b}. Our analysis finds
FPP\,=\,0.025\%, independently demonstrating (with high likelihood)
that the candidate is a planet.

\subsubsection{Previously Identified Candidates}
\label{sec:prev}
Several planet candidates showing just a single transit each were
discovered in \ktwo Campaigns 1--3 \citep{osborn:2016}. Since our
transit search focuses on shorter-period planets (see
Sec.~\ref{sec:vetting}), we did not identify these systems.

K2-44 (EPIC~201295312) was identified as hosting a planet candidate by
\cite{montet:2015} and Doppler spectroscopy constrains its mass to be
$<12 M_\oplus$ \citep[95\% confidence;][]{vaneylen:2016}.  Our
analysis of this system yields FPP\,$<$\,\minfpp and so validates this
previously identified candidate.

Of the 9 planet candidates identified by \cite{montet:2015}, we
validate five as planets: K2-44, K2-46, K2-8, K2-27, K2-35 (EPIC
201295312, 201403446, 201445392, 201546283, and 201549860,
respectively).  For three candidates (EPIC~201702477, 201617985, and
201565013), we find \minfpp$<$\,FPP\,$<$\,\maxfpp. For EPIC 201828749
we find FPP$<$\minfpp, but a nearby star seen via high-resolution
imaging prevents us from validating this candidate.
 
The largest single sample of \ktwo planet candidates released to date
are the 234 candidates identified by \cite{vanderburg:2016} in
Campaigns 0--3.  Our analysis independently identifies 127 of their
candidates. Of these 127, we validate 72 as planets and identify 19 as
false positives. Our analysis validates several multi-planet candidate
systems announced in that work: K2-23, K2-58, K2-59, K2-62, K2-63, and
K2-75 (EPIC 206103150, 206026904, 206027655, 206096602, 206101302, and
206348688, respectively).

Furthermore, our analysis identifies 69 new candidates not published
in the sample of \cite{vanderburg:2016}; these are mostly in Campaign 4, some are in earlier
Campaigns.  The two samples largely overlap, but each also contains
many candidates identified by only a single team.  The differences
between the two samples (along with our non-detection of
EPIC~206318379, noted above) suggests that multiple independent
analyses are essential if many planet candidates are not to be missed.

When comparing our sample with that of \cite{vanderburg:2016}, we find
that the largest single systematic difference between is that they
report roughly 25\% more candidates with $P<5$\,d. Our vetting checks
suggest that most of these excess short-period planets are eclipsing
binaries.  In particular, our early-stage vetting procedures
(described in Sec.~\ref{sec:vetting}) indicate that EPIC~201182911,
201270176, 201407812, 201488365, 201569483, 201649426, 202072965,
202086968, 202093020, 202843107, 203942067, 204649811, 205463986, and
206532093 are all likely false positives. Furthermore, high-resolution
imaging of a random selection of four of their candidate systems
revealed all four to have nearby multiple stars (EPIC 203099398,
203867512, 204057095, 204750116). While these newly-detected stars do
not prove that the systems are false positives, they will nonetheless
complicate any effort to  validate these candidates.

Aside from the apparent excess of short-period false positives in the
\cite{vanderburg:2016} sample, we find no statistical differences
between the properties of their and our candidate samples.
Measurements of both pipelines' detection efficiencies could determine
why each team has missed so many of the candidates detected by the
other group. The implication for future surveys is that multiple
independent pipelines may substantially increase the total survey
completeness of independent, relatively low-budget survey programs.

\cite{adams:2016} report nine new candidates in Campaigns 0--5 with
$P<1$\,d. Of their five new candidates in Campaigns 0--4, we identify
and validate one: K2-85b (EPIC~210707130b), which hosts a small planet
on a 16\,hr period.  Because our transit search did not extend to
ultra-short orbital periods, we did not identify EPIC 202094740,
203533312, 210605073, or 210961508.



\cite{schmitt:2016} identify several dozen systems as likely eclipsing
binaries (see their Table 1). Of these we identify four:
EPIC~201324549 is a false positive while EPIC~201626686, 204129699,
and 206135267 remain candidate planets.  Of their planet candidates we
find three to have low FPPs EPIC~201920032, 206061524, and 206247743)
and we validate five as planets (K2-55, K2-60, K2-67, K2-73, and K2-76
-- respectively: EPIC 205924614, 206038483, 206155547, 206245553, and
206432863).  We did not identify their candidate EPIC~201516974
because of its 36.7\,d orbital period.


%


\begin{figure}[ht!]
\begin{center}
\includegraphics[width=3.1in]{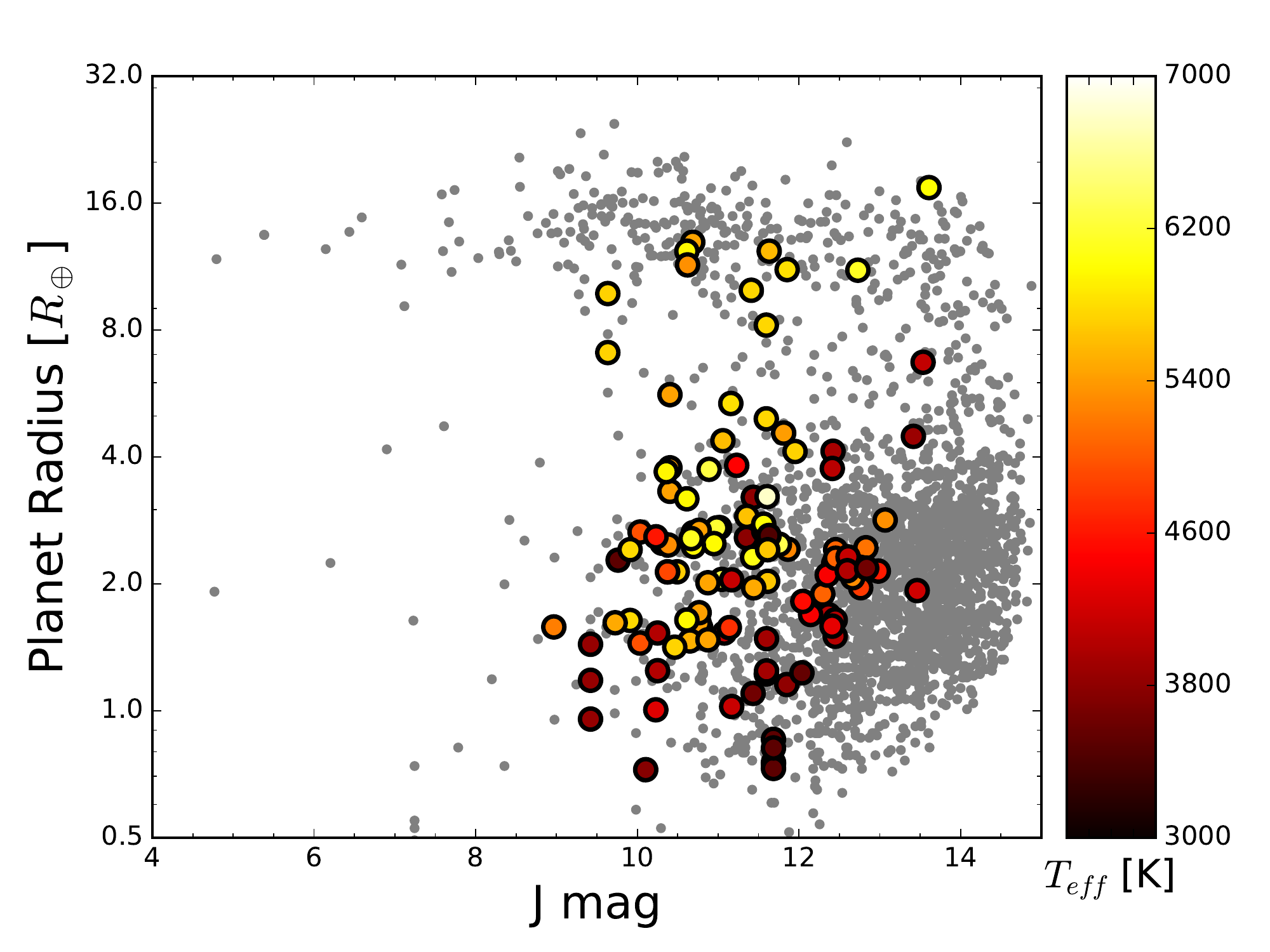}
\hspace{0.5in}
\includegraphics[width=2.7in]{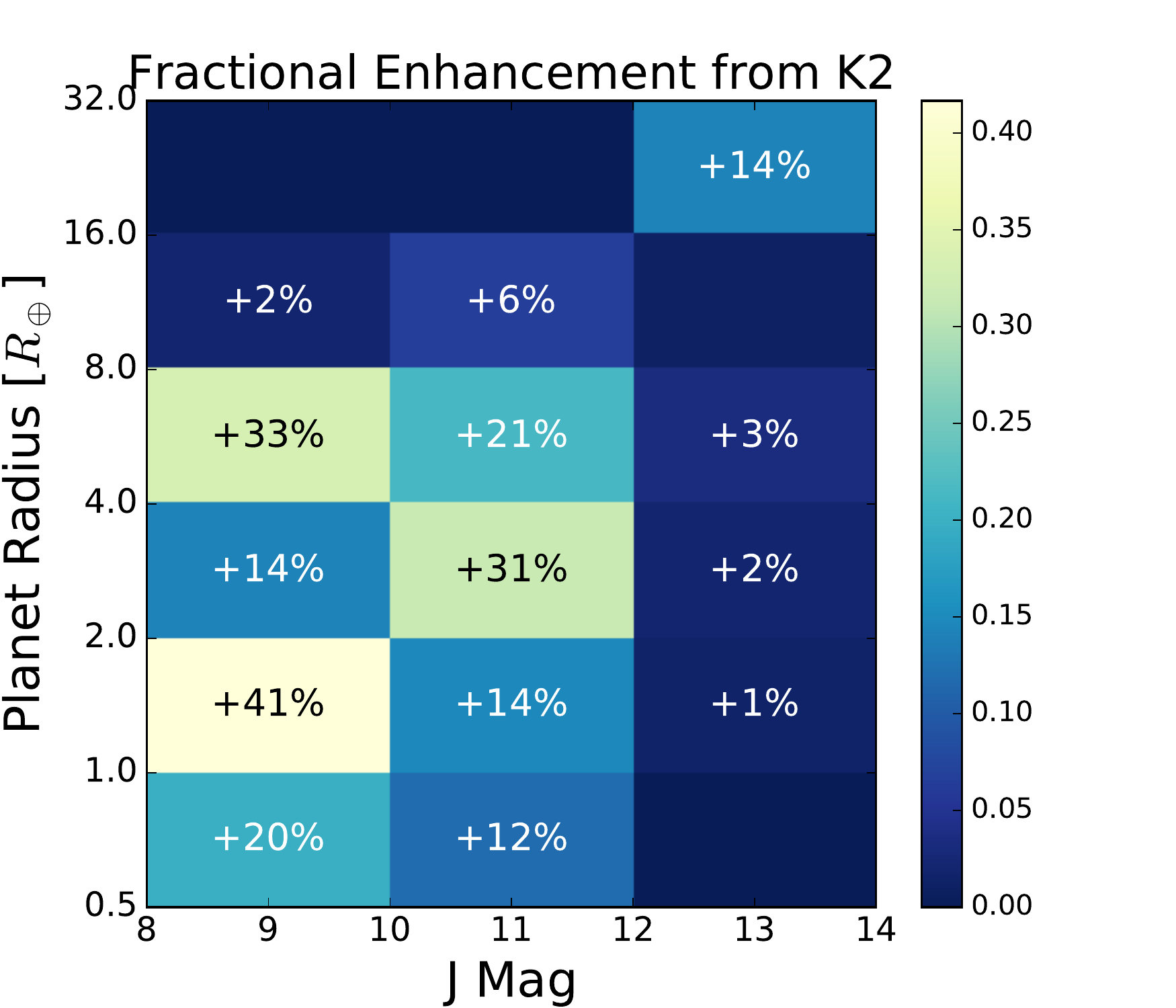}
\caption{\label{fig:jmag_radius_teff} {\em Left:} Planetary radius,
  stellar magnitude, and \Teff for all validated planets (colored
  points) and all planets at the NASA Exoplanet Archive (gray points).
  {\em Right:} Fractional enhancement by \ktwo to the population of
  known planets.  In its first five fields, \ktwo has already
  substantially boosted the numbers of small, bright planets.}
\end{center}
\end{figure}

\section{Conclusion and Final Thoughts}
\label{sec:conclusion}

We have presented \nplanet validated planets discovered using \ktwo
photometry and supporting ground-based observations. Of these,
\newplanet are planets validated here for the first time. Our analysis
shows that \ktwo has increased by 30\% the number of small
(1--4\,$R_\oplus$) planets orbiting bright stars ($J=8-12$\,mag), as
depicted in Fig.~\ref{fig:jmag_radius_teff}. We report several new
multi-planet systems, including the four-planet system K2-72 (EPIC 206209135);
for all these systems we verify that the derived stellar parameters
are consistent for each planet in each system. Our analysis finds
\nfinalcand remaining planet candidates, which likely include a
substantial number of planets waiting to be validated.  In this work,
we specifically utilize all our available follow-up data to assess the
candidate systems.  We claim to validate candidates only when no other
plausible explanations are available; for example, many systems remain
candidates because of nearby stars detected by our high-resolution
imaging.

The size of our validated-planet sample demonstrates yet again the
power of high-precision time-series photometry to discover large
numbers of new planets, even when obtained from the wobbly platform of
\ktwo.  Since \ktwo represents a natural transition from the
narrow-field, long-baseline \Kepler mission to the nearly all-sky,
mostly short-baseline \tess survey, the results of our \ktwo efforts
bode well for the productivity of the upcoming \tess mission. The
substantial numbers of intermediate-sized planets orbiting moderately
bright stars discovered by our (and other) \ktwo surveys
(Fig.~\ref{fig:jmag_radius_teff}) will be of considerable interest for
future follow-up characterization via radial velocity spectroscopy and
JWST transit observations \citep[e.g.,][]{greene:2016}.

We searched the entire sample of \ktwo targets without regard for the
different criteria used to propose all these stars as targets for the
\ktwo mission. Thus although the planet population we present is
broadly consistent with the early candidate population discovered by
\Kepler \citep[e.g.,][]{borucki:2011}, our results should not be used
to draw conclusions about the intrinsic frequency with which various
types of planets occur around different stars. To do so, we are
already investigating a full end-to-end measurement of our survey
completeness as a function of planet and stellar properties. By doing
so, we also hope to compare the quality of the various input
catalogs and selection metrics used to pick \ktwo targets.

Both \ktwo and \tess offer the potential for exciting new demographic
studies of planets and their host stars. \ktwo observes a
qualitatively different stellar population than \Kepler, namely a much
larger fraction of late-type stars \citep{huber:2016}. Stellar
parameters for these late-type systems derived from photometry alone
are relatively uncertain, and follow-up spectroscopy is underway to
characterize these stars (Dressing et al., in prep; Martinez et
al., in prep). In addition to the difference in median spectral type,
\ktwo also surveys a much broader range of Galactic environments than
was observed in the main \Kepler mission.  These two factors suggest
that, once \ktwo's detection efficiency is improved and quantified, the
mission's data could address new questions about the intrinsic
frequency of planets around these different stellar populations.

At present, when comparing our planets and candidates with those
identified by \cite{vanderburg:2016} we find only a partial
overlap between the two samples. This result could imply significant,
qualitative differences in vetting effectiveness and survey
completeness, and suggests that the analysis of transit survey data by
multiple teams is an essential component of any strategy to maximize
the number of discoveries.  As noted in Sec.~\ref{sec:fpp}, we
estimate that our sample has an overall false positive rate of
15--30\% (depending on the FPR of candidates with additional nearby
stars), with an indication that FPP increases for larger sizes and
shorter periods.

We therefore re-emphasize that lists of \ktwo candidates and/or validated
planets are not currently suitable for the studies of planetary
demographics that \Kepler so successfully enabled
\citep[e.g.,][]{howard:2012,mulders:2015b}. The best path forward to
enabling such studies would seem to include robust characterization of
pipeline detection efficiency, as was done with \Kepler
\citep{petigura:2013a,dressing:2015,christiansen:2015}. It may be that
such an approach, combined with further refinement of the existing
photometry and transit detection pipelines, would allow the first
characterization of the frequency of planet occurrence with Galactic
environment across the diverse stellar populations probed by \ktwo's
ecliptic survey.

Barring unexpected technical mishaps, \ktwo is currently capable of
operating through at least C18. The number of targets observed in
these first \ktwo campaigns contain comparable numbers of targets to
later campaigns (with the exception of Campaign 0, which had a
duration only roughly half that of the later, $\sim$80-day campaigns).
If \ktwo continues to observe, based on current discoveries we would
expect a planet yield roughly 4--5 times as great as that currently
produced. Accounting for the relatively large survey incompleteness
revealed by comparing our results to other \ktwo surveys
\citep{vanderburg:2016}, we expect \ktwo to find anywhere from
500--1000 planets over its total mission lifetime. Analysis and
follow-up of these systems will occupy exoplanet observers up to the
\tess era, and beyond.


\vspace{1in}


\acknowledgements

For useful suggestions and comments on an early draft we thank Andrew
Vanderburg, Knicole Colon, Tim Morton, and Michael Werner.  We thank
Katherine de Kleer and Imke de Pater for contributing observations,
and our many, many telescope operators and observing assistants for
helping us obtain such a wealth of data.

This paper includes data collected by the K2 mission. Funding for the
K2 mission is provided by the NASA Science Mission directorate.

This work was performed in part under contract with the Jet Propulsion
Laboratory (JPL) funded by NASA through the Sagan Fellowship Program
executed by the NASA Exoplanet Science Institute.  E.A.P. acknowledges
support by NASA through a Hubble Fellowship grant awarded by the Space
Telescope Science Institute, which is operated by the Association of
Universities for Research in Astronomy, Inc., for NASA, under contract
NAS 5-26555.  The Research of J.E.S was supported by an appointment to
the NASA Postdoctoral Program at the NASA Ames Research Center,
administered by Universities Space Research Association under contract
with NASA. B.J.F. was supported by the National Science Foundation
Graduate Research Fellowship under grant No. 2014184874. Any opinion,
findings, and conclusions or recommendations expressed in this
material are those of the authors and do not necessarily reflect the
views of the National Science Foundation.  A.J. acknowledges support
from FONDECYT project 1130857, BASAL CATA PFB-06, and from the
Ministry for the Economy, Development, and Tourism's Programa
Iniciativa Cient\'{i}fica Milenio through grant IC\,120009, awarded to
the Millennium Institute of Astrophysics (MAS).

Some of the data presented herein were obtained at the W.M. Keck
Observatory (which is operated as a scientific partnership among
Caltech, UC, and NASA) and at the Infrared Telescope Facility (IRTF,
operated by UH under NASA contract NNH14CK55B). The authors wish to
recognize and acknowledge the very significant cultural role and
reverence that the summit of Maunakea has always had within the
indigenous Hawaiian community.  We are most fortunate to have the
opportunity to conduct observations from this mountain.

C.B. acknowledges support from the Alfred P. Sloan Foundation. The
Robo-AO system was developed by collaborating partner institutions,
the California Institute of Technology and the Inter-University Centre
for Astronomy and Astrophysics, and with the support of the National
Science Foundation under Grant Nos. AST-0906060, AST-0960343, and
AST-1207891, the Mt. Cuba Astronomical Foundation and by a gift from
Samuel Oschin.

The Pan-STARRS1 Surveys (PS1) have been made possible through
contributions of the Institute for Astronomy, the University of
Hawaii, the Pan-STARRS Project Office, the Max-Planck Society and its
participating institutes, the Max Planck Institute for Astronomy,
Heidelberg and the Max Planck Institute for Extraterrestrial Physics,
Garching, The Johns Hopkins University, Durham University, the
University of Edinburgh, Queen's University Belfast, the
Harvard-Smithsonian Center for Astrophysics, the Las Cumbres
Observatory Global Telescope Network Incorporated, the National
Central University of Taiwan, the Space Telescope Science Institute,
the National Aeronautics and Space Administration under Grant
No. NNX08AR22G issued through the Planetary Science Division of the
NASA Science Mission Directorate, the National Science Foundation
under Grant No. AST-1238877, the University of Maryland, and Eotvos
Lorand University (ELTE).

{\it Facility:} \facility{APF (Levy)}, \facility{\Kepler}, \facility{\ktwo}, \facility{Keck-I (HIRES)}, \facility{Keck-II (NIRC2)}, \facility{IRTF (SpeX)}, \facility{Palomar:Hale (PALM-3000/PHARO)}, \facility{Palomar:1.5m (Robo-AO)}, \facility{Gemini North (DSSI)}, \facility{Gemini South (NIRI)}, \facility{LBT (LMIRCam)}

\input{reamatch_results_v4.tex}
\input{feros_rvs.tex}
\input{aocomp_all_v7.tex}
\input{aoval_v6.tex}
\input{stellar_results_v7.tex}

\clearpage
\begin{landscape}
\input{candidate_results_v8.tex}
\clearpage
\end{landscape}

\input{nonplanet_candidate_results_v6.tex}
\input{HRI_observations_v2.tex}


\clearpage
\bibliographystyle{apj_hyperref}

\input{ms.bbl}

\end{document}

%% file: newcommand_gen.tex
\newcommand{\rsun}{$R_{\odot}$\xspace}

\newcommand{\rearth}{$R_\earth$\xspace}

%% file: fprate_results_v6.tex
\begin{deluxetable}{ll}
\tabletypesize{\scriptsize}\tablecaption{False Positive Rates}\tablehead{
                   Category & FP Rate \\
}\startdata
       $R_P \le 2 R_\oplus$ &    0.07 \\
 $2 \le R_P/R_\oplus \le 8$ &    0.08 \\
       $R_P \ge 8 R_\oplus$ &    0.54 \\
               $P \le 3$\,d &    0.36 \\
           $3 \le P \le 15$ &    0.12 \\
              $P \ge 15$\,d &    0.21 \\
              Entire Sample &    0.20 \\
\enddata\label{tab:fprates}
\end{deluxetable}

%% file: reamatch_results_v4.tex
\begin{deluxetable}{rll}
\tabletypesize{\scriptsize}\tablecaption{HIRES Follow-up Observations}\tablehead{
      EPIC & Flag &                                               Note \\
}\startdata
  201295312 &    1 &                                                 1\% \\
  201338508 &    1 &                                                 1\% \\
  201367065 &    1 &                                                 1\% \\
  201384232 &    1 &                                                 1\% \\
  201393098 &    1 &                                                 1\% \\
  201403446 &    1 &                                                 1\% \\
  201445392 &    1 &                                                 1\% \\
  201465501 &    1 &                               1\% \\
\enddata\tablenotetext{1}{No detection of second spectrum at noted flux ratio.}
\tablenotetext{2}{Star is unfit for ReaMatch: Teff below 3500 K.}
\tablenotetext{3}{Star is unfit for ReaMatch: Vsini above 10\,km~s$^{-1}$.}
\tablenotetext{4}{Ambiguous detection.}
\tablenotetext{5}{Obvious detection}\label{tab:reamatch}
\end{deluxetable}

%% file: feros_rvs.tex
\begin{deluxetable}{l l r r r r}
\tabletypesize{\footnotesize}
\tablecaption{FEROS Radial Velocities}
\tablehead{EPIC     & BJD & RV & $\sigma_{RV}$ & $t_{int}$ & S/N}
 \startdata
203294831 &    2457182.56854355 &   6.235  & 0.093  &   600  &   45\\
203771098 &    2457182.57744262 &   0.713  & 0.010  &   600  &   82\\
201176672 &    2457182.58920619 &  41.702  & 0.029  &  1800  &   25\\
201626686 &    2457182.62251241 &  49.133  & 0.015  &   600  &   49\\
203929178 &    2457182.67319105 &-130.200  & 0.221  &   600  &   54\\
203771098 &    2457183.50141524 &   0.725  & 0.010  &   600  &   78\\
203929178 &    2457183.53736243 &  64.251  & 0.276  &   600  &   51\\
201626686 &    2457183.54133676 &  49.090  & 0.011  &   600  &   76\\
\enddata                                                           
\label{tab:ferosrv}                                                
\end{deluxetable}                                                  

%% file: aocomp_all_v7.tex
\begin{deluxetable}{rlrrll}
\tabletypesize{\scriptsize}\tablecaption{HRI-Detected Stellar Companions}\tablehead{
      EPIC &  $r_{ap}$ &  $\rho$ &  Delta mag & Filter &   Telescope \\
           &  [''] &  [''] &  [mag] &  &    \\
}\startdata
 201176672 &    11.94 &   0.340 &       2.77 &      K &       Keck2 \\
 201295312 &    11.94 &   8.110 &       4.00 &      K &       Keck2 \\
 201295312 &    11.94 &   8.070 &       4.10 &      K &     Palomar \\
 201324549 &    11.94 &   0.090 &       0.43 &      K &   GemN-NIRI \\
 201488365 &      --- &   4.100 &       9.56 &      i &     Robo-AO \\
 201505350 &    11.94 &   0.160 &       0.36 &      K &     Palomar \\
 201546283 &     7.96 &   3.030 &       5.83 &      i &     Robo-AO \\
 201546283 &     7.96 &   2.960 &       3.74 &      K &       Keck2 \\
\enddata\label{tab:aocomp}
\end{deluxetable}

%% file: aoval_v6.tex
\begin{deluxetable}{rlll}
\tabletypesize{\scriptsize}\tablecaption{Disposition of Multi-star Candidates}\tablehead{
      EPIC & $\rho<4$" & $F_2/F_1<\delta'$ &                             Comment \\
}\startdata
 201176672 &      True &              True &          Cannot validate candidate. \\
 201295312 &     False &              True &          Same depth for $r=1$'' aperture.\\
 201324549 &      True &              True &          Cannot validate candidate. \\
 201546283 &      True &              True &          Cannot validate candidate. \\
 201626686 &     False &              True &          Shallower transit with $r=1$''; likely FP.\\
 201629650 &      True &              True &          Cannot validate candidate. \\
 201828749 &      True &              True &          Cannot validate candidate. \\
 201862715 &      True &              True &          Cannot validate candidate. \\
\enddata\label{tab:aoval}
\end{deluxetable}

%% file: stellar_results_v7.tex
\begin{deluxetable}{rrlllll}
\tabletypesize{\scriptsize}\tablecaption{Stellar Parameters}\tablehead{
  EPIC &      Kp &       $R_*$ &       $M_*$ &    $T_{eff}$ &     log $g$ & Source\\
           &    [mag]&  [$R_\odot$] &      [$M_\odot$] &    [K]  & [dex] & \\
}\startdata
 201155177 &  14.632 &   0.643(39) &   0.702(46) &     4613(71) &   4.659(50) &   Huber et al.\ (2016)\\
 201176672 &  13.980 &   0.508(98) &   0.559(87) &    4542(130) &   4.747(97) &   Huber et al.\ (2016)\\
 201205469 &  14.887 &   0.570(30) &   0.600(30) &     3939(87) &   4.698(23) &   Huber et al.\ (2016)\\
 201208431 &  14.409 &   0.435(60) &   0.487(72) &     4044(81) &   4.849(60) &   Huber et al.\ (2016)\\
 201247497 &  16.770 &   0.436(27) &   0.492(29) &     3918(46) &   4.846(50) &   Huber et al.\ (2016)\\
 201295312 &  12.126 &    1.58(15) &   1.150(60) &     5912(51) &   4.101(63) &   SpecMatch\\
 201324549 &  12.146 &    1.45(25) &    1.18(12) &    6283(113) &    4.17(12) &   Huber et al.\ (2016)\\
 201338508 &  14.364 &   0.462(38) &   0.520(44) &     4021(62) &   4.823(50) &   Huber et al.\ (2016)\\
\enddata\label{tab:stars}
\end{deluxetable}

%% file: candidate_results_v8.tex
\begin{deluxetable}{lllllllllll}
\tabletypesize{\scriptsize}\tablecaption{Planet Candidate Parameters}\tablehead{
Other  &     EPIC &            $P$ &          $T_{0}$ &    $T_{14}$ &   $R_P/R_*$ &          $a$ &      $R_P$ &  $S_{inc}$ &      FPP &                                 Disposition \\
Name &       &             [d] &          [BJD$_\textrm{TDB}$  &     [hr] &   [\%] &          [AU] &      [$R_\oplus$] &    [$S_{\oplus}$] &       &                                  \\
& & & -- 2454833] 
}\startdata
K2-42b & 201155177.01 &    6.68796(93) &    1981.6763(52) &    3.04(20) &    3.04(28) &   0.0617(13) &   2.15(24) &  44.1(6.3) &  5.2e-07 &                                      Planet \\
 & 201176672.01 &    79.9999(98) &    2044.8046(15) &   11.75(17) &   18.0(1.1) &    0.299(16) &  10.2(2.1) &   1.10(46) &    0.094 &       Candidate (see Tables~\ref{tab:aoval} and \ref{tab:feros}) \\
K2-43b & 201205469.01 &    3.47114(21) &    1976.8845(33) &    1.97(11) &    6.60(36) &  0.03784(63) &   4.13(31) &  49.0(6.9) &  9.7e-10 &                                      Planet \\
K2-4b & 201208431.01 &    10.0044(11) &    1982.5170(45) &    2.87(17) &    3.49(41) &   0.0715(35) &   1.69(30) &   8.9(2.7) &  0.00028 &                                      Planet \\
 & 201247497.01 &    2.75391(15) &    1977.9031(23) &    1.03(14) &    7.9(1.3) &  0.03035(59) &   3.78(68) &  43.7(6.1) &     0.41 &                                   Candidate \\
K2-44b & 201295312.01 &    5.65688(59) &    1978.7176(44) &    4.36(13) &    1.56(12) &   0.0651(11) &   2.72(32) &    646(126) &  6.7e-07 &                                      Planet \\
 & 201324549.01 &   2.519334(35) &   1976.99353(69) &   1.545(15) &    2.41(30) &   0.0383(13) &   3.91(85) &   2002(727) &        1 &  False Positive (see Tables~\ref{tab:aoval} and~\ref{tab:reamatch}) \\
K2-5c & 201338508.01 &    10.9324(14) &    1981.6012(55) &    2.84(23) &    3.24(29) &   0.0775(22) &   1.64(20) &   8.3(1.6) &  2.8e-05 &                                      Planet \\
\enddata\label{tab:results}
\end{deluxetable}

%% file: nonplanet_candidate_results_v6.tex
\begin{deluxetable}{lllllllll}
\tabletypesize{\scriptsize}\tablecaption{Unvalidated Candidate False Positive Likelihoods}\tablehead{
       Target &    L\_heb\tablenotemark{a} & L\_heb\_Px2\tablenotemark{a} &     L\_eb\tablenotemark{b} & L\_eb\_Px2\tablenotemark{b} &    L\_beb\tablenotemark{c} & L\_beb\_Px2\tablenotemark{c} &     L\_pl\tablenotemark{d} &      FPP \\
}\startdata
 201176672.01 &   3.2e-36 &        0.0097 &  7.3e-19 &        0.046 &     0.002 &        0.0019 &    0.035 &    0.094 \\
 201247497.01 &   7.2e-16 &       7.4e-05 &      2.6 &           20 &         0 &           2.9 &       12 &     0.41 \\
 201445392.01 &   0.00027 &         0.028 &    0.042 &         0.83 &         0 &         0.081 &      1.5 &    0.042 \\
 201465501.01 &   0.00047 &          0.19 &   0.0036 &         0.46 &         0 &         0.015 &    0.042 &     0.45 \\
 201512465.01 &    0.0021 &          0.09 &    0.013 &         0.82 &       1.3 &           5.6 &      8.3 &     0.23 \\
 201546283.01 &     6e-15 &         1e-07 &  8.1e-10 &      5.4e-16 &    0.0019 &       0.00045 &       14 &  4.4e-07 \\
 201565013.01 &   3.6e-07 &         0.073 &       16 &          1.2 &       1.1 &          0.44 &        8 &     0.76 \\
 201617985.01 &    0.0048 &          0.77 &   0.0046 &          1.5 &      0.09 &             0 &    0.025 &     0.96 \\
\enddata\label{tab:fppbreakdown}
\tablenotetext{a}{\ Likelihood that system is a heirarchical eclipsing
  binary, with orbital period either as measured or twice that measured.}
\tablenotetext{b}{\ Likelihood that system is an eclipsing
  binary, with orbital period either as measured or twice that measured.}
\tablenotetext{c}{\ Likelihood that system is a blended eclipsing
  binary, with orbital period either as measured or twice that measured.}
\tablenotetext{d}{\ Likelihood that system is a transiting planet.}
\end{deluxetable}

%% file: HRI_observations_v2.tex
\begin{deluxetable}{lrrl}
\tabletypesize{\scriptsize}
\tablecaption{High-Resolution Imaging}
\tablehead{
EPIC & Filter & $t_{int}$ (s) & Instrument
}\startdata
 201155177 &      K &    330 &       NIRI \\        
 201176672 &      K &    270   &      NIRC2 \\      
 201205469 &      K &    810   &      NIRC2 \\      
 201208431 &      K &    171   &      NIRC2 \\      
 201247497 &      K &    540   &      NIRC2 \\      
 201295312 &      K &    212.4 &      PHARO \\      
 201295312 &      K &    225   &      NIRC2 \\      
 201324549 &      K &    276.1 &      PHARO \\
\enddata
\label{tab:hri}
\end{deluxetable}